%% file: RISsurveyIEEE.tex
\documentclass[10pt,twocolumn]{IEEEtran}

\usepackage{graphicx,epsfig}
\usepackage{amssymb}
\usepackage{amsmath}
\usepackage{multicol}
\usepackage[figuresright]{rotating}
\usepackage{bm}
\usepackage{caption}
\usepackage{lipsum}
\usepackage{subfigure}
\usepackage{epsfig}
\usepackage{multirow}
\usepackage{amssymb}
\usepackage{amsmath}

\usepackage{array}

\usepackage{verbatim}
\begin{document}
%
\title{Reconfigurable Intelligent Surface: Design the Channel  -- a New Opportunity for Future Wireless Networks}


\author{Miguel~Dajer, Zhengxiang~Ma, Leonard~Piazzi, Narayan~Prasad, \\
         Xiao-Feng Qi, Baoling Sheen, Jin Yang, and Guosen~Yue$^*$ 

\thanks{
 The authors are with Wireless Access Lab, Futurewei Technologies, Inc., Bridgewater, NJ 08807, USA.  
Emails: \{mdajer, zma, lpiazzi, nprasad, xqi, bsheen, jyang, gyue\}@futurewei.com }
\thanks{\noindent$^*$ Corresponding author.}
}



%


\maketitle
\date{}
\begin{abstract}

In this paper, we survey state-of-the-art research outcomes in the burgeoning field of reconfigurable intelligent surface (RIS) in view of its potential for significant performance enhancement for next generation wireless communication networks by means of adapting the propagation environment. Emphasis has been placed on several aspects gating the commercially viability of a future network deployment. Comprehensive summaries are provided for practical hardware design considerations and broad implications of artificial intelligence techniques, so are in-depth outlooks on salient aspects of system models, use cases, and physical layer optimization techniques.

\end{abstract}








%

\input{01-Intro}

\input{02-SysModel}

\input{03-PhyDesign}

\input{04-AntRF_IEEE}

\input{05-AIML}

\input{06-Challenges}

\input{07-Conclusions}




%




\bibliographystyle{IEEEtran}

\bibliography{RIS_IEEE}

\end{document}

%% file: 01-Intro.tex
\section{Introduction}\label{sec.intro}

As the world witnesses growing deployments of the fifth generation (5G) network, the wireless industry is casting its eyes on beyond-5G~(B5G), or 6G technological breakthroughs, in anticipation of proliferating new applications and use cases that demand a wireless communication-and-sensing network with vastly higher agility, coverage and throughput \cite{DoCoMo2020WhitePaper, Saad2020Network, Oulu2019}. Emerging B5G applications include autonomous driving, tactile remote interaction, and augmented reality, to name a few. New technological enablers capable of delivering a 10x performance gain over their 5G counterparts will be required to fulfill such applications, much like how massive multi-input-multi-output (MIMO), emerging as a promising research subject a decade ago, became a critical enabler for the evolution of commercial deployment from 4G to 5G.

Among candidate technologies at a nascent research stage, reconfigurable intelligent surfaces (RIS) have received much interest in academia and industry \cite{DiRenzo20JSAC, DocomoMetawave2018}. A wireless mobile communication network traditionally adapts transmission schemes of the transceiver end-points to mitigate or leverage the dynamic multipath propagation environment for optimal performance, but has not had the means to control the environment itself. An RIS offers a potential solution to achieve a software configurable smart radio environment. Conceptually, it is a large and thin metasurface of metallic or dielectric material, comprised of an array of passive sub-wavelength scattering elements with specially designed physical structure, the elements can be controlled in a software-defined manner to change the electromagnetic (EM) properties (e.g. phase shift) of the reflection of the incident radio frequency (RF) signals. By a joint phase control of all scattering elements, the reflected radiation pattern of the incident RF signals can be arbitrarily tuned in real time, thus creating new of degrees of freedom to the optimization of the overall wireless network performance. The phase-only version of the RIS is often referred to as intelligent reflecting surface (IRS) \cite{DiRenzo20JSAC}.

The passive and conformal nature of an RIS should allow for straight-forward integration on existing surfaces (e.g. walls, furniture), non-invasively augmenting the environment of existing wireless networks for increased spectral efficiency (SE) and energy efficiency (EE), without altering the wireless standards and designs for transceivers already in deployment. In addition to enhancing the performance of wireless communications networks, RIS has been identified as aiding sensing, localization, and wireless power transfer (see, e.g. \cite{vacarubio2020primer, SHu18TSP, Mishra2019ICASSP}, among others). In particular, it is potentially beneficial for networks operating in high frequency bands (e.g. millimeter-wave, terahertz) where blockage- and absorption-induced dead zones and channel rank deficiency are pervasive performance bottleneck that would otherwise necessitate ultra-dense base station (BS) deployment with increased power consumption and backhaul cost (e.g. \cite{Tan2018INFOCOM, Abari17mmNets,Mustafa:BlockRIS}).

Several review articles on RIS have been published \cite{Basar19Access,QWu20ComMag,yuan2020RIS}. More recent ones \cite{DiRenzo20JSAC,SGong20CommST} provide comprehensive, state-of-the-art accounts of surface and network models, ultimate limits on network performance, optimal design at the electromagnetic, physical and network layers, comparisons against legacy technologies (massive MIMO, active relay), outstanding challenges and future directions of this burgeoning field. In particular, \cite{DiRenzo20JSAC} emphasizes the importance of adopting physically consistent surface interaction models, based on synthesis and analysis tools derived from surface electromagnetic (SEM) theory. A common theme is that system level understanding of the ultimate performance limit of an RIS network involving both active endpoints and distributed passive surfaces, and the network architecture responsible for such performance, is still in its infancy \cite{QWu20ComMag,DiRenzo20JSAC,SGong20CommST}. It is apparent that the progress made so far has been primarily achieved through means of analysis and simulation, often on systems under rather ideal assumptions, e.g. perfectly known environment state information, infinite-resolution phase control, simplified electromagnetic model of the surface, while also overlooking protocol and computation overhead associated with a real-world network deployment. Furthermore, since the propagation environment has a much larger physical footprint than the communications end-points, power concentration on receivers, formerly achieved by co-located coherent endpoint beamforming (e.g. massive MIMO) must now be achieved by multiple distributed passive surfaces over a considerably larger surrounding in order to capture sufficient transmitted energy that can then be manipulated by the surfaces. To justify a commercially viable deployment, unit-area cost and power consumption must be sufficiently low for the total cost of ownership to be on par with that of a conventional co-located or distributed MIMO network with similar performance. Therefore, an inexpensive low-power surface is critical to its successful commercial deployment, where tailored physical and network layer optimization techniques are required to deliver better performance at lower cost and power consumption. A more comprehensive understanding of recent experimental studies to this end (see, e.g. \cite{Arun20RFocus,ZLi19GetMobile}) is therefore warranted. On a separate note, machine learning (ML) and artificial intelligence (AI) techniques have been covered in the aforementioned review papers, albeit in a brief and somewhat ad hoc manner, as a promising augmentation to model-based approaches in RIS system level optimization especially for massive deployment of inexpensive surfaces that defy parametric modeling.

This review paper aims to accentuate individual RIS research aspects deemed relevant to a \emph{system level} tradeoff between performance, cost and power consumption, a prerequisite for a viable real-world network deployment. Hence, besides the basic concept of RIS, channel model, and the physical layer (PHY) design optimizations, we provide an in-depth review on hardware design and implementation, and a comprehensive summary of ML and AI techniques applied to RIS system. The rest of the paper is organized as follows. Section~\ref{sec.sysmodel} summarizes practical system level IRS use cases and corresponding channel models; Section~\ref{sec.phydesign} highlights studies on physical layer designs and algorithms taking into account design constraints resulting from system performance-cost-power tradeoffs; Section~\ref{sec.antrf} elucidates the hardware needs, reconfigurable nature, complexities, and future directions, linking today's RIS efforts to the learned experiences from traditional transmitarray and reflectarray antenna research; Section~\ref{sec.aiml} highlights recent works on applying ML and AI techniques to optimization of real-world RIS networks that defy simple analytical characterization; Section~\ref{sec.challenges} articulates system level open questions facing a viable real-world deployment of RIS-augmented wireless networks; Section~\ref{sec.con} concludes the article.
\\
\\ 

%% file: 02-SysModel.tex
\section{RIS System and Channel Model}\label{sec.sysmodel}
\subsection{Reconfigurable surface and basic functionalities}\label{sec.sysmodel.basicfunc}
\begin{figure}[h]
\center{ \centerline{ {\epsfxsize=2in\epsffile{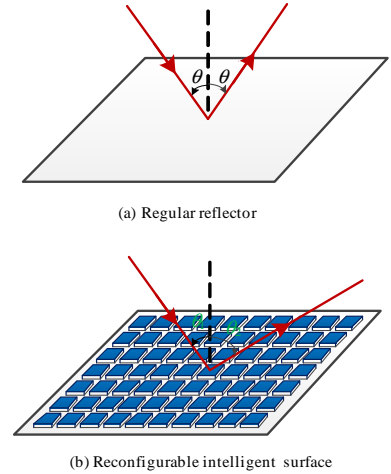}}}
\caption{ Reconfigurable surface} \label{fig.RISvsRegS}}
\end{figure}

Fundamentally, when a wireless signal, or in general, an EM wave reaches the boundary between two isotropic mediums, the relationship between the angle of incidence and the angle of reflection and refraction is governed by Snell's law.  Particularly for the reflection, for regular surface of the medium, the angle of incidence is the same as the angle of reflection as shown in Figure~\ref{fig.RISvsRegS}(a). Recent research advances on the reflectarrays with metasurface make  it possible to change the impedance of the surface and achieve a certain phase shift between the incident and scattered waves~\cite{Hum14TAP}. When the surface is divided into a large number of  closely-spaced elements and each metasurface element is made to have appropriate phase shifts, the EM wave is tuned to some other angle instead of the symmetric reflective wave based on  Snell's law. Ideally, as shown in~Figure~\ref{fig.RISvsRegS}(b), if the phase shift of each metasurface element can be configured to any value, a reflected beam at any angle can be formed~\cite{Ozdogan20WCL}. However, to form the reflected beam with incident EM wave, the phase shift of the surface element has to be set appropriately or smartly, where signal processing design techniques or machine learning based approaches can be applied, resulting in the so-called reconfigurable \emph{intelligent} surface. In practice, continuous or analog phase shifts may be difficult to achieve.  Recently, as reported in~\cite{Hum14TAP,LZhang18NatureCom,LDai20Access},  the configurable RIS's with quantized phase shifts are designed.

\begin{figure}
  \center{\includegraphics[width=.95\linewidth]{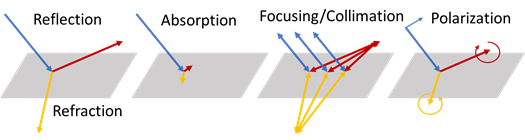}}
  \caption{Illustration of basic functionalities of an RIS. Adapted from~\cite{DiRenzo20JSAC}.}
  \label{fig.RISbasicfun}
\end{figure}

While the most research interests of RIS are shown on the intelligent reflecting surface (IRS) as described above,  other functionalities of RIS are also worth more research efforts. As illustrated in~Figure~\ref{fig.RISbasicfun}, if we limit ourselves to situations where either the input or output signal is a plane wave, the basic functionalities of RIS can be summarized as the following~\cite{DiRenzo20JSAC}: 1) reflection/refraction, where a plane wave is diverted from its original direction of propagation to another direction; 2) absorption, where a plane wave is substantially reduced in its amplitude; 3) focusing/collimation, where a plane wave is focused to a single point or a spherical wave from a point source is converted to a single plane wave; 4) polarization modification, where the operation 1) or 3) also involves changing the polarization of the incoming wave, such as from linear polarizations to circular polarizations.  
There is virtually no difference between RIS and the more traditional reflectarray/transmitarray antennas~\cite{Berry63TAP,KLam97APMConf} as far as these basic functionalities are concerned. In addition, reconfigurability requires that the parameters of each functionality can be dynamically changed, for example, the direction of reflection/refraction, the focal point position, etc.

\subsection{RIS Systems}
\begin{figure}[h]
\center{ \centerline{ {\epsfxsize=2.8in\epsffile{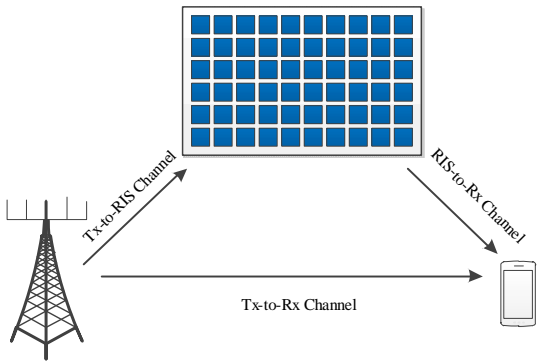}}}
\caption{ A wireless transceiver system assisted with a reconfigurable intelligent surface} \label{fig.RISsystem}}
\end{figure}

As shown in Figure~\ref{fig.RISsystem}, a basic RIS system for research studies consists of a transmitter, a receiver, and an RIS panel with programmable phase shifts and/or the reflective amplitude. The RIS panel reflects the  incident signal from the transmitter. From the perspective of the wireless communication system, with controllable phase shift on each element, the RIS changes the channel environment from the transmitter to the receiver.  The resulting  channel between the transmitter and receiver now consists of two components: the direct channel from the transmitter to the receiver without involving the RIS and the channel with the RIS interaction. With appropriate design on the RIS phase shifts or amplitude or both, a certain metric objective such as the system achievable rate or the coverage can be optimized by changing the channel environment. This is fundamentally different from conventional wireless communication research where the design and optimization opportunities are confined to the pair of transceivers.

\subsubsection{RIS Configurability}
As we can see, the key feature that makes the RIS attractive is its ability to change the wireless environment, or specifically, the wireless channel between the transmitter and receiver to improve the system performance. We present a brief summary of the configurability of the RIS based on hardware designs as follows.

\smallskip\noindent{\emph{On-Off RIS:}} A simple configuration of RIS is a one-bit on-off switch, which has been designed and built in~\cite{Arun20RFocus}.  Each surface element can be configured in one of two states, i.e., either the ``on" state when the signal is reflected or the ``off" state when the signal passes through the element. There is no configurable phase shift on the incident signal for reflection in the design. It is claimed that with a properly designed method of on-off selection, the RIS is able to achieve the power gain in the order of $N^2$ where $N$ is the number of elements in the RIS~\cite{Arun20RFocus}. However, with ideal on-off RIS, such 1-bit on-off RIS without any phase-shift, there is limited potential  to reshape the reflected beam pattern for the far-field incident radio wave. The power of the reflected signal that reaches the user location may be significantly lower than the total power of the reflected signal. Another design of on-off RIS can be found in~\cite{BZHu10APL}, where the surface material can be switched between nearly total reflection and total absorption of a particular polarized incident wave through appropriate bias voltages that  turn on and off diodes in the RIS design.

\smallskip\noindent{\emph{Phase-Shift RIS:}} In the phase-shift RIS,  a phase shift is introduced on the impinging radio wave so that the reflected radio signal over all the elements can be controlled towards a desired location or coherent signal combining at the desired receiver~\cite{Hum14TAP}. The reflection amplitudes can be different among the surface elements but they cannot be actively controlled. Currently, most research studies on the RIS focus on the phase-shift only variant, including system design~\cite{taha2019enabling,CHuang19SPAWC,huang2020reconfigurable} and hardware design~\cite{LDai20Access,YZhou20SciRep}.

\smallskip\noindent{\emph{Phase and Amplitude Control:}} As the title suggests, in this RIS category, both phase shift and reflection amplitude can be controlled. As in~\cite{LZhang18NatureCom}, the so-called digital metasurface is a designed reflective surface loaded with positive-intrinsic-negative (PIN) diodes. By applying control voltages on the PIN diode, the reflection coefficients of the surface elements can be dynamically controlled with discrete phase or amplitude states.  With more controlled variables, more design flexibilities can be exploited. Consequently, superior performance can be achieved. The disadvantage is the significant complexity increase. Another scenario in this category is the amplitude is controlled by the phase shift. As shown in~\cite{Abeywickrama20TCom},  the reflection amplitude coefficient of the surface element is a function of configured phase-shift. In this case, although the amplitude cannot be independently configured, it can be designed through the phase consideration by jointly considering the impacts on the amplitude and phase-shift over the channel. If the model in~\cite{Abeywickrama20TCom} holds for most practical hardware design even with the phase shift, the joint consideration on both the phase shift and the amplitude is important for future studies.

A special case of the third category is a combination of the first two. If the hardware design allows, for the phase-shift RIS, an ``off" state can be introduced to each element. Rather than always reflecting the incident signal with a phase-shift, the element can be switched to the ``off" state  and let the signal pass through as in~\cite{Arun20RFocus}. In this case the amplitude is set to 0 when it is ``off" and the fixed scalar when a phase-shift is configured. This is certainly an interesting design perspective for future research, which could have a large impact on the resulting performance, particularly  for the current hardware designs with very limited discrete phase shifts~\cite{LDai20Access}.

\subsubsection{System Settings}

We now describe several system settings for RIS PHY research based on  different antenna and user configurations. 
From the PHY design considerations, different settings in the following may lead to different design objectives and philosophies.

\smallskip\noindent{$\bullet\;$   Single antenna transceiver systems }

\smallskip In RIS assisted wireless communication with single antenna at the transmit and receiver ends,  the channel between the RIS and the transmitter and the reflected channel after RIS interactions are both vectors. The effective channel from transmitter to receiver for the transmitter-RIS-receiver path is a scalar, i.e., a single-input single-output (SISO) channel. With single antenna at the transmitter, the transmission design does not impact the spatial channel characteristics at the RIS. The design is focused on the RIS configurations. The SISO case is a system setting for studying the RIS behavior and impact on the overall channel. It is also a common system setting for indoor RIS design. Many RIS research works start with emphasis of single antenna  systems as in~
\cite{Basar19EuCNC,CYou20JSAC,taha2019enabling,YYang20TCom}, where \cite{Basar19EuCNC} derives the error probabilities  of the RIS-assisted SISO transceiver system, \cite{CYou20JSAC} provides the design of discrete phase values on the RIS to maximize the achievable rate, \cite{taha2019enabling} and \cite{YYang20TCom} considers wideband OFDM transmissions and design phase shifts for the wideband transmission with different channels over different subcarriers.

\smallskip\noindent{$\bullet\;$  MIMO systems}

\smallskip With multiple antennas at the transmitter and receiver, MIMO has been widely adopted in commercial wireless systems. Here we include the cases of the single receive antenna, i.e., multiple-input single-output (MISO), since the transmit beamforming exploiting channel spatial diversity is one key feature of MIMO systems (other than spatial multiplexing). Since the transmit precoding or beamforming changes spatial characteristics of the signal, the interaction with RIS is then different. The RIS design needs to be adapted to the particular transmit precoding. Since the transmit precoding can be very dynamic, e.g., changing over every time slot, joint design of RIS and transmit precoding is essential.

\smallskip\noindent{\emph{Single user MIMO:}}
For single user MIMO, the spatial multiplexing and/or spatial diversity of the MIMO channel can be exploited. These two aspects are also studied in the RIS-assisted single user MIMO systems, i.e., how to improve the beamforming gain with the channel path through a large RIS or the spatial multiplexing by enriching the channel scattering.
In \cite{QWu18GC} joint design of transmit active beamforming and passive RIS beamforming is considered to exploit channel diversity with RIS for a downlink MISO system.
In \cite{Ozdogan20ICASSP}, the authors consider the RIS phase shift design to improve the rank of the channel matrix, i.e., the spatial multiplexing gain and, consequently, the MIMO capacity.
The study in \cite{QWu20JSAC} considers the joint transmit and RIS beamforming design for the RIS assisted simultaneous wireless information and power transfer (SWIPT) with design objective of  transmit power minimization.

\smallskip\noindent{\emph{Multi-user MIMO (MU-MIMO):}}
In MU-MIMO, particularly for massive MIMO, the transmitter with an antenna array can serve multiple users simultaneously by exploiting multiuser spatial diversity with multiuser beamforming/precoding. For RIS assisted MU-MIMO, it is clear that with multiple RIS's, the spatial diversity among users can be improved. However,  with a single RIS panel, designing an efficient RIS to improve MU-MIMO performance is challenging as the transmit-RIS path is the same for all users, particularly for rank-deficient line-of-sight (LoS) channel between the transmitter and the RIS. Therefore joint design of RIS and transmit MU precoding is essential.   Here are several studies on how the RIS improves MU-MIMO performance.

In~\cite{CHuang19TWC}, considering the zero-forcing (ZF) precoding at the transmitter for composite channels with RIS phase shift, the RIS design for MU-MISO is formulated as a joint optimization of both the transmit power allocation and the phase shifts of the surface reflecting elements. The problem is extended to wideband OFDM system in~\cite{HLi20WCNC}, where the transmit beamformer
and the RIS phase shifts are jointly designed to maximize the average sum-rate over all subcarriers. \cite{MZhao20WCL} considers symbol level precoding for RIS-aided MU-MISO to minimize symbol error probability with QAM and PSK inputs.
Both studies in \cite{GZhou20WCL, zhang2020robust} consider the robust beamforming for RIS-assisted MU-MIMO transmissions  while the later considers a cognitive radio system and optimizes secondary transmit beamforming  subject to a limited interference imposed on the primary user. In \cite{RLiu20WCNC} joint symbol level precoding and reflection coefficients design is considered for RIS-assisted MU-MISO systems to minimize the transmit power with guaranteed QoS.

It is worth noting that investigations into multiple RIS assisted single and multi-user MIMO communications have also been initiated. In particular, theoretical analysis of using two surfaces to establish a double reflection link for single-user communications is pursued in \cite{Han:DoubleIRS}, where it is shown that under  specific geometrical channel model and configuration, a scaling in received SNR of $O(N^4)$, where $N$ is the total number of elements across the two RISs, can be possible.
On the other hand, \cite{Mustafa:BlockRIS} considers a network with several blockers. It employs  stochastic geometry based tools to quantify the  impact of coating a fraction of blockers with configurable meta-surfaces on several system metrics (such as probability of being in a blind spot etc.).

It is noted that here we aim to categorize pertinent system settings for future research, as well as point out a few exemplary existing works for further reading. A most recent and comprehensive summary on the existing research works for different system settings described above can be found in~\cite{SGong20CommST}.

\subsection{Channel Model}
\subsubsection{RIS Interaction}
A simple channel model on the RIS interaction has been considered in many research studies. Consider signal antenna at both the transmitter and the receiver. Assume the number of elements at the RIS is $N$.  Denote the channel between the transmitter and RIS and between RIS and receiver as $\mathbf{h}_{Tx,RIS}$ of size $N\times1$ and $\mathbf{h}_{RIS,Rx}$, respectively. Denote $\phi_i$ and $c_i$, $i=1,...,N$, as the phase shift and the scaling factor  at the RIS element $i$. The overall channel   from the transmitter to the receiver  is given as~\cite{YYang20TCom,wu2020intelligent}
\begin{equation}
h_{RIS}=\mathbf{h}_{Tx,RIS}^T \mathbf{\Phi} \mathbf{h}_{RIS,Rx}\label{eq.h_SISO_RIS}
\end{equation}
where $\mathbf{\Phi}$ is a diagonal matrix with the entry $\mathbf{\Phi}_{i,i}=c_i e^{j\phi_i}$. In most studies, it is assumed that $c_i=1$ for all elements, i.e., full signal reflection~\cite{Taha19GC,Taha20SPAWC,taha2019enabling,Elbir20WCL,CHuang19TWC,CHuang20JSAC}. The model can be easily extended to the case of multiple antennas at the transmitter and the receiver.

More recently, a new RIS channel model is proposed in~\cite{Abeywickrama20TCom} where it is claimed that the proposed new model is more practical. In this model, instead of a constant scaling factor $c_i$ for each surface element, it is modelled as a function of the phase shift $\phi_i$, i.e., $c_i(\phi_i)$, given as
\begin{equation}
 c_i(\phi_i)=(1-c_{\min})\left(\frac{\sin(\phi_i-\phi_0)}{2}\right)^k+c_{\min},
\end{equation}
where $c_{\min}$, $\phi_0$ and $k$ denote the minimum amplitude, the horizontal distance
between $-\pi/2$ and $c_{\min}$, and the steepness factor of the
function $c_i(\phi_i)$, respectively. All three parameters are non-negative constants determined by the specific design and implementation.

\subsubsection{Channel Model for Transmitter-RIS-Receiver Link}

\noindent{\bf  Pathloss model: } Assuming the far-field approximation, the large-scale pathloss for the transmitter-RIS-receiver is the product of pathlosses of the transmitter-RIS and RIS-receiver links. The average received power, $P_{ RIS}$ at the receiver for the transmitter-RIS-receiver link follows a production model which is inversely proportional to the product of the  transmitter-RIS and RIS-receiver distances~\cite{tang2019wireless,wu2020intelligent}, given by
\begin{equation}
P_{RIS}\propto \frac{1}{d_{ Tx-RIS}^{a_{Tx-RIS}}  d_{RIS-Rx}^{a_{RIS-Rx}}},
\end{equation}
where $d_{Tx-RIS}$ and $d_{RIS-Rx}$ denote the distances between the transmitter and the RIS, and the RIS and the receiver, respectively, $a_{Tx-RIS}$ and $a_{RIS-Rx}$ are corresponding pathloss exponents.

For more practical considerations, the pathloss models as well as the shadowing in the specifications of the 3rd Generation Partnership Project (3GPP) can be considered. Depending on the deployments of the RIS systems, e.g., indoor or outdoor, urban or rural, high-rise building or on the ground,  the pathloss and shadowing models for the transmitter-RIS and RIS-receiver links can be selected accordingly from~\cite{3GPPTR36.814V9,3GPPTR38.901}.

\smallskip\noindent{\bf Channel model for fading:}
Appropriate models for small-scale fading channels $\mathbf{h}_{Tx,RIS}$ and $\mathbf{h}_{RIS,Rx}$ are desirable for RIS research efforts. For conventional MIMO, i.i.d. Rayleigh fading model has been widely considered in the literatures which is fine for a limited number of widely spaced antennas. While in some research works, the study and design of RIS is still based on the i.i.d. fading channel~\cite{CHuang20JSAC}, for the RIS with closely spaced elements, such a fading model may not be appropriate.  The design methodologies in these papers may still be valid. However, one needs to be cautious on the results as well as the conclusions drawn from these results, particularly when designing the experimental prototype and the practical systems, since the channel will never be i.i.d. in a real-world RIS  environment.

For appropriate RIS channel modeling, it is necessary to consider the spatial channel model (SCM). One simple spatial channel model is multi-scattering ring model for multiple channel clusters as described in~\cite{yue19GJSDMwcnc} for one-dimensional array. Given an angle spread $[-\Delta_l,\Delta_l]$ and the power distribution $\rho_l(\theta)$ for scattering ring $l$, the spatial correlation for each scattering ring $\mathbf{R}_l$ can be obtained with each entry given as \cite{yue14Access,yue20JSAC}
\begin{equation}
{[\mathbf{R}_l]}_{i,j} = \int_{-\Delta_l}^{\Delta_l}\rho_l(\alpha)e^{j\mathbf{k}^T(\alpha+\theta_l)(\mathbf{u}_i-\mathbf{u}_j)}d\alpha.\label{eq.R_l}
\end{equation}
where $\theta_l$ is center of angle of arrival (AoA) of the cluster $l$,  $\mathbf{k}=\frac{-2\pi}{\lambda}(\cos(\alpha),\sin(\alpha))^T$ is the wave number vector for a planar wave with angle of arrival $\alpha$, $\lambda$ is the carrier wavelength, $\mathbf{u}_i$ and $\mathbf{u}_j$ are the position  vectors of element $i$ and $j$, respectively. The channel power distribution $\rho_l(\theta)$ can be uniform, truncated Laplace, or truncated Gaussian distribution. The overall spatial covariance matrix for multi-scattering rings is a linear combination of $\mathbf{R}_l$, i.e., $\mathbf{R}=\sum_{l}\tilde{\rho}_l \mathbf{R}_l$, where the weight $\tilde{\rho}_l$ is the power ratio for scattering ring $l$ and $\sum_{l}\tilde{\rho}_l=1$~\cite{yue19GJSDMwcnc}. Replaced with 3D angles in (\ref{eq.R_l}), the multi-ring model can be easy extended to 3D channel model with a 2D array for RIS systems.

Based on eigen-decomposition of covariance matrix $\mathbf{R}$,  the fading channel can be generated using the Karhunen-Loeve representation with Gaussian complex random vectors. For correlated MIMO channels, we obtain the covariance matrices for both ends, $\mathbf{R}^{t}$ and $\mathbf{R}^r$, with the multi-ring model and generate the MIMO channel as in~\cite{Goldsmith03JSAC}.

For wideband OFDM systems, the channel delay profile can be introduced in an extended scattering ring model as in~\cite{yue20HiDiConvGC} where each scattering ring is assigned with a path delay. With one scattering ring assigned with zero delay which represents the channel cluster for the first arriving signal path, other path delays are uniformly distributed in a region of $[0,\tau_{DS}]$ where $\tau_{DS}$ is the delay spread.  With  multi-scattering rings and a delay profile, a spatial-frequency covariance is defined in~\cite{yue20HiDiConvGC} for vectorized channel in both spatial and frequency domains. The wideband channel can be generated based on the specified spatial-frequency covariance, again using the Karhunen-Loeve transformation.

The aforementioned scattering ring model is a simple far-field statistical channel model, which is sufficient for modelling the channels for the transmitter-RIS link and RIS-receiver link.  Based on the corresponding discrete spatial channel model with power angle profile, the MIMO fading channel for 1-D antenna array can be generated as $\mathbf{H}= \sqrt{\frac{1}{L}} \sum_{l=1}^{L} h_{l}\mathbf{b}_{r}(\theta_{r,l}) \mathbf{b}_{t}^H(\theta_{t,l})$, where $L$ is the number of paths, $ \mathbf{b}_t(\theta_{t,l})$ and $ \mathbf{b}_{r}(\theta_{r,l})$ are the beam steering vectors for the transmit and receive end  with their entries given as $b_{t,i}(\theta)=e^{j\mathbf{k}^T(\theta)\mathbf{u}_{t,i}}$ and $b_{r,i}(\theta)=e^{j\mathbf{k}^T(\theta)\mathbf{u}_{r,i}}$, respectively, and $h_{l}$ with $\mathbb{E}\{|h_l|^2\}=p_l$ is complex channel gain on each path. The power profile $\{p_l\}$ can be generated according to a certain distribution. For some deployment scenarios, particularly, the Tx-RIS channel, the channel can be considered sparse, meaning that the number of paths or clusters is small and for the multi-ring model the angle spread is also small. For the outdoor high-rise RIS scenario, the channel can be modeled as single-path line-of-sight.
For more practical channel models, particularly for system level evaluations,  3D  channel model specified in 3GPP report can be used for RIS studies for different deployment scenarios~\cite{3GPPTR38.901,3GPPTR36.873}. Given a specific 3-D environment map, a ray-tracing based channel model can also be considered~\cite{taha2019enabling,RemcomRayTracing}.

\smallskip\noindent{\emph{Near-field Channel Modeling:}}
The far-field assumption may not hold for the scenarios in which the RIS is sufficiently large or close to either transmitter or receiver, particularly for the indoor environment with a large RIS, e.g., the settings in~\cite{Arun20RFocus}. Therefore, in these cases, we need to take into account near-field propagation properties in the channel modeling~\cite{yuan2020RIS}. However, there are few studies that consider the near-field scenario  in the literature. In~\cite{tang2019wireless}, the  free-space pathloss model for RIS-assisted communications is developed. A general pathloss formula is provided that can be applied for the near-field scenario. It is shown that the free-space pathloss in the near field is proportional to $(d_{Tx-RIS}+d_{RIS-Rx})^2$ instead of the product $d_{Tx-RIS}^2d_{RIS-Rx}^2$ for the far-field case. Based on the general pathloss formula, the pathloss formulas for near-field beamforming and broadcasting are derived to characterize the free-space path loss of RIS-assisted beamforming and broadcasting. Experiments in a microwave anechoic chamber are conducted and the measurement results match well with the modelling results.
Most recent studies and analysis on near-field channel for MIMO and RIS-assisted communications can be found in~\cite{Bjornson20OJCom,Pizzo20JSAC}.

\subsection{Challenges and Future Studies}
A significant challenge for the modeling efforts described in this section is the channel model with the RIS interaction. Almost all the existing research on the physical layer design for RIS assumes perfect phase shift modeling for the composite transmitter-RIS-receiver channel after RIS interaction. Therefore,  the accuracy of the phase shift model on the impinging radio signal is critical. The phase-shift model error will greatly impact the design performance.  As in most RIS design on the phase shift, an optimization problem is formulated and solved. With the modeling error, the solution may no longer be optimal. If it is difficult to model the precise phase shift, an error model on the phase shift needs to be studied. Such an error model can be used to evaluate the design algorithms in terms of performance sensitivity of the solutions with respect to the modeling assumptions.

The modeling error issue becomes more critical when the authors of~\cite{Abeywickrama20TCom} show that the amplitude of RIS reflection is a function of phase shift and provide a model on the scaling factor. The proposed model is verified with the numerical results with the design parameters from experiments in~\cite{ZhuActiveImp}. However, it is still a question as to how well the model matches practical hardware implementations.  It is then a question of how the error impacts the system design when using the exact mathematical expression in the model as in design formulations.  Nonetheless, it may be safe to use the model in~\cite{Abeywickrama20TCom} for numerical evaluations.

Another key aspect in IRS modeling is to incorporate the coupling between its closely spaced elements. While a model that captures coupling with high fidelity is clearly beneficial,  tractability of that model will determine how conducive it is for optimizing different criteria such as energy efficiency and spectral efficiency.
 In this context, tools developed in \cite{KunduThesis} for modeling compact MIMO arrays can be a useful starting point at least in the far-field regime.

Better modeling of the RIS interactions  is an important direction for future research,  particularly for various hardware design of RIS described later in Section~\ref{sec.antrf}. On the other hand, if it is difficult to reduce the modeling error, it is then worthy to investigate the robust design in future studies.

As aforementioned, near-field channel modeling with RIS interactions is less studied. In particular for RIS deployed in the indoor environments, at home or in enterprise settings, near-field channel modeling is another important research area for future studies.

%% file: 03-PhyDesign.tex
\section{Physical Layer Design and Algorithms}\label{sec.phydesign}
\subsection{Channel Estimation and Reconstruction}
It is evident that  without the availability of reliable channel estimates the phase patterns (or RIS response) cannot be optimized. However,  channel estimation in RIS aided communications  is quite challenging in practice and is elucidated in this section.
Since almost all of the works on channel estimation consider time-division duplexing (TDD), we will begin by considering RIS assisted TDD systems and later briefly consider the extension to frequency-division duplexing (FDD) systems.
Clearly, the key advantage in TDD systems is the channel reciprocity. However, there is a penalty to pay in terms of latency and coverage in the uplink. The latter  arises from having to switch in time between uplink and downlink, and as a result of forsaking the use of a lower carrier frequency in the uplink thereby incurring a higher path-loss. In addition, for RIS assisted communications there is yet another classification. In particular, we can classify explicit channel estimation schemes into two broad categories. The first category is formed by those requiring the use of a few active elements and receive RF chains or receive baseband processing capability at the  RIS, whereas the second category of schemes assumes purely passive RIS elements with no such baseband processing capability. Let us consider these two categories in more detail. 

\smallskip\noindent $\bullet\;$ {\em Channel estimation with baseband processing capability at the RIS} 
 
\smallskip\noindent In schemes that fall in this category each of the channel links in (\ref{eq.h_SISO_RIS}), namely   $\mathbf{h}_{Tx,RIS}$ and $\mathbf{h}_{RIS,Rx}$  are {\em separately} estimated. As a result, the rich toolkit of techniques developed for massive MIMO channel estimation can be readily applied, especially those that seek to reconstruct the full channel relying on a limited number of analog observations, each observation being a projection of that channel corrupted by noise.
Indeed, by  exploiting reciprocity in TDD, each of these channel vectors can be estimated by transmitting  pilots from the transmitter and receiver, respectively, and collecting the corresponding analog observations at the RIS. An estimate (or reconstruction) of each channel vector is obtained by processing these collected analog observations. In particular,  \cite{taha2019enabling}
reconstructs each channel vector by relying on compressive sensing tools. Indeed, each of the
 channel vectors $\mathbf{h}_{Tx,RIS}$ and $\mathbf{h}_{RIS,Rx}$ are assumed to have a sparse representation in the beam domain which is then exploited via sparse reconstruction formulations. On the other hand,
 \cite{MatcompGC} proposes an  Alternating Direction Method of Multipliers (ADMM) based algorithm by adopting a matrix completion based formulation. In addition to sparsity, the formulation in \cite{MatcompGC}  also exploits a low-rank property of an underlying matrix and models the available observations  as randomly sampled entries of that matrix that are further corrupted by noise.  

\smallskip\noindent$\bullet\;$ {\em Channel estimation without baseband processing capability at the RIS}

\smallskip\noindent In this category of schemes the RIS is assumed to have all passive elements.
An  early representative work  under this category, \cite{He:CasChEst}, proposes a two-stage scheme for (single-user) channel estimation. Here the first stage utilizes approximate message passing techniques for sparse bilinear matrix factorization to obtain the RIS-to-user channel along with another matrix term. In the second stage the latter matrix term is used as the input to a low-rank matrix completion algorithm to obtain the RIS-to-TX channel. The caveat is that this two-stage scheme
assumes that each RIS element can be turned on and off and in its on state its reflection coefficient can be configured to any arbitrary phase term.
Another representative example is the 3-stage scheme proposed in \cite{ChEstWang} which derives dimensioning rules to decide the number of pilot resources for a given configuration comprising of the number of RIS, BS antennas as well as the number of users and  antennas per-user. While \cite{ChEstWang} assumes a worst-case modeling in which no sparsity is assumed for any channel links and all channels are supposed to be full rank, a key observation that  the RIS-TX link is common for all users is nevertheless exploited. This allows for a much more benign scaling of pilot overhead in the number of users.
Yet another notable work is
 \cite{chen2019channel} which again proposes a multi-stage scheme. This works exploits the aforementioned fact that  the RIS-TX link is common for all users. In addition, supposing a uniform linear array at the RIS, it also
 exploits
 a particular row-column block sparsity structure possessed by the effective channel in RIS aided communications, that does not arise in conventional massive MIMO systems.

A key fact of  \cite{ChEstWang,chen2019channel,He:CasChEst} is that channel estimation is initiated from scratch in that no {\it a priori} side-information which can reduce overhead is assumed to be available.
 In this context, we note that exploiting subspace side-information together with other reconstruction methods can very significantly reduce the training overhead, as demonstrated by the recent results in \cite{Prasad:TDDirs}.
Moreover, akin to machine learning based approaches, a location indexed dictionary can be maintained which can  provide such side-information (cf. \cite{CHuang19SPAWC}) or at-least a good initial point (aka hot start) to subsequent refinement.
Constructing and maintaining such a dictionary for RIS assisted communications is an interesting avenue for future research,  where we also remark that a key advantage of subspace estimation is that it needs to be done at a coarser time-scale compared to instantaneous channel estimation.
 We further note that effective algorithms  for subspace estimation from low dimensional projections over massive MIMO systems have been proposed in \cite{Haghi:LowD}.  One attempt to tailor these algorithms  to RIS-assisted communications has been made in \cite{Prasad:TDDirs}, wherein a particular type of inherent sparsity is exploited.

The   works discussed so far have all considered TDD wherein analog observations are available as inputs to the reconstruction algorithm. To the best of our knowledge, one of the very few works  directly applicable to an FDD scenario  where instantaneous channel reciprocity cannot be exploited or more broadly to a scenario where quantized feedback information must be employed as input to the reconstruction algorithm,  is  \cite{Arun20RFocus}. In fact, \cite{Arun20RFocus} directly estimates an optimized RIS pattern by relying  on average received signal strength   feedback. However, while \cite{Arun20RFocus} assumes no simplifying structure in the underlying cascaded channel, the voting based algorithm suggested there is applicable to only on-off control of RIS elements. Indeed, directly determining an optimized  RIS on-off pattern instead of reconstructing the underlying effective
channel is justified in \cite{Arun20RFocus} to be more sample efficient by invoking the limited on-off control of RIS elements that is possible.  More recently, a multiple beam training scheme has been proposed in \cite{You:BeamTraining} albeit after adopting a simplified geometrical channel model. This scheme allows users to determine their respective optimized beams and can significantly reduce overhead while offering similar performance compared to conventional beam training schemes that entail a full sweep over available beam directions.
On the other hand,  given channel links having non-LoS components and where finer control of RIS elements is possible, explicit channel reconstruction becomes important as we seek to realize the gains achievable via an optimized RIS pattern. For such reconstruction, an interesting avenue for research would be to extend the non-convex quadratically constrained quadratic programming based FDD massive MIMO channel reconstruction
  proposed in \cite{PrasadWiOpt19}, which combines subspace side-information with limited instantaneous quantized feedback information.
 In this context, we note that dictionary based interpolation approaches for inferring downlink covariance (and hence downlink subspace information) based on observed uplink estimates have been gainfully exploited over conventional massive MIMO communications \cite{DecurningeCov}.

Another key aspect in training over RIS-assisted communications is the design of RIS patterns suitable for the training phase. While \cite{chen2019channel} proposes certain formulation for designing RIS patterns based on an incoherence criterion that facilitates sparsity based reconstruction, 
 design of RIS patterns that can be effectively used during the training phase is an important open problem.  We recall that   generalized Lloyd algorithms  have traditionally been used for vector quantizer design and more recently for massive MIMO analog codebook design \cite{Ganji:AnCB}.
Such  algorithms can also be leveraged to construct a set of RIS patterns suitable for training, where
finite alphabet constraints on these pattern vectors can be imposed via a proximal distance based approach (see \cite{Prasad:TDDirs} for one such provably convergent algorithm).

In summary, while channel reconstruction (estimation) for a single-RIS assisted communication has developed rapidly,   an effective multi-time scale framework to estimate covariance (subspace) and instantaneous channel coefficients  has not yet been developed. Such a framework will entail effective provisioning of training resources, reconstruction algorithms, as well as tracking and change detection algorithms.
The design challenges as well as potential gains will be substantially magnified as we move onto multi surface scenarios.  One of the very few practical works that have leveraged multiple re-configurable surfaces is \cite{Abari17NSDI}. However, the scenario in \cite{Abari17NSDI} is single-user communications in which multiple surfaces provide a selection diversity in that at each instant a single surface is chosen (based on location and directional information) and configured to assist communications.

\subsection{Optimizing RIS-assisted Communications}
Multiple works have designed algorithms for   throughput gains, coverage improvements, enhancing energy efficiency and optimizing other metrics  over   RIS-assisted communication systems, given accurate channel state information.
A key motivation is the $O(N^2)$ rate of improvement in received signal power that can be achieved by increasing the number $N$ of RIS elements \cite{WuIRS}. 
We add the caveat here that this  rate of growth does not hold for arbitrarily large $N$ and breaks down as we transition to the near field regime, but nevertheless it can be accrued for large   RIS configurations of practical interest \cite{Bjornson20OJCom}.
Notable examples of RIS optimization are \cite{CHuang19TWC} wherein energy efficiency advantages of RIS over amplify-forward relaying are demonstrated and \cite{wu_discrete_tc} wherein a joint beamforming and RIS interaction matrix optimization is considered under the restriction that the RIS elements can be assigned discrete phase shifts.

More recently, non-ideal RIS elements that entail phase-dependent amplitude attenuation have been modeled in \cite{Abeywickrama20TCom} by leveraging previous modeling results and experimental validation provided in \cite{ZhuActiveImp}.
Design algorithms matched to this  non-ideal model have been proposed there and it is shown that even single-user SNR (or effective channel power gain) maximization problem which has a simple coherent combining and maximal phase alignment based optimal solution in the case of RIS with ideal elements, can become non-trivial under a non-ideal element model. The latter non-ideality  introduces a  tradeoff between ensuring phase alignment and reducing amplitude attenuation.   Penalty-based algorithms are suggested in \cite{Abeywickrama20TCom} to address single and multiuser joint beamforming and pattern optimization problems. The key message there is that non-idealities must be explicitly incorporated in the problem formulation, and naive approaches which directly apply solutions derived assuming ideal RIS elements can incur substantial performance degradation.

We remark that the  design optimization considered in all hitherto mentioned works have assumed a model with uncoupled RIS elements. This assumption is particularly convenient for optimization since it allows for controlling the RIS response on a per-element basis. However, it can only be justified for an RIS with sufficiently spaced apart elements (a.k.a. Nyquist spacing).  A relevant problem that arises   for a given RIS aperture size, is to determine the quantum of additional gains that can be achieved by closely packing a larger number of  elements as opposed to using  a fewer number of elements with Nyquist spacing. 
To assess these gains, a  proper modeling of compact RIS  (along with potentially compact transmit and receive arrays) followed by design of well matched optimization algorithms are imperative.  

%% file: 04-AntRF_IEEE.tex
\section{Review of RIS Hardware Design}\label{sec.antrf}
\subsection{Introduction}

In this section, we discuss the hardware design aspects of RIS. The RIS envisioned is an electrically thin (thickness $\ll$ wavelength) 2D structure with a large footprint (width/length $\gg$ wavelength). It can be planer or conformal with curvatures. When we limit ourselves to consider a small area (a few wavelengths) of the RIS that is nearly flat, and ignore the long range interactions between different parts of a large RIS, there is no differnece between RIS and the more traditional reflectarray/transmitarray antennas~\cite{Berry63TAP,KLam97APMConf,NayeriReflectAntBook,Abdelrahman17SynLectAnt} in terms of the EM functionalities they can provide, as summarized earlier in Section~\ref{sec.sysmodel.basicfunc}. However, since RIS is envisioned to be used to modify the radio environment, and in the real world, the wavefront is usually more complex than a single plane wave due to the complex multi-path propagation, there is additional requirement on RIS.

Obviously, any complex wavefront may be decomposed into a linear combination of plane waves of various directions and complex amplitudes. To deal with such complex wavefront, ideally, the RIS should support each of the functionalities with multiple parameters simultaneously. For example, to focus a complex wavefront to a single point requires the RIS to focus each of the component plane waves with different incoming directions to the same focal point coherently. Or for an even more sophisticated example, in a multi-user situation, an RIS may be required to simultaneously focus multiple complex wavefronts to multiple focal points corresponding to the location of the multiple users. Actually, simple examples of such complex requirements have been implemented in reflectarray/transmitarray~\cite{Nayeri12TAP,Nayeri13APSURSI}, except for the reconfigurability. In~\cite{Nayeri12TAP}, a reflectarray antenna is designed and prototyped, and supports 4 beams simultaneously from a single left-hand circular polarized (LHCP) feed horn. In~\cite{Nayeri13APSURSI}, a quadrufocal transmitarray antenna is designed and prototyped, and supports 4 focal points simultaneously.

In the rest of this section, we begin with conceptual architectures to realize the basic functionalities of RIS using nearly passive RF hardware, and examples of actual RIS implementations based on those architecture. We then provide an overview of reflectarray and transmitarray antennas, which are the predecessors to RIS as far as RF hardware design is concerned, and typical technologies for realizing reconfigurable reflectarray and transmitarray antennas. We conclude with RIS hardware design challenges and potential future research directions.

\subsection{Conceptual Architectures of RIS}

\begin{figure}
  \center{\includegraphics[width=.98\linewidth]{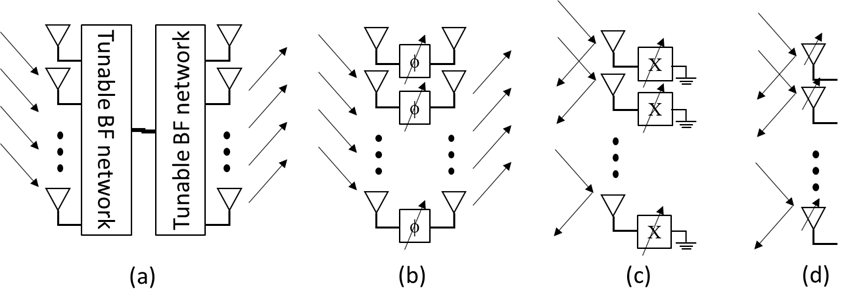}}
  \caption{Conceptual architectures of RIS with array of antenna elements.}
  \label{fig.RISconceptarch}
\end{figure}

\begin{figure}
  \center{\includegraphics[width=.98\linewidth]{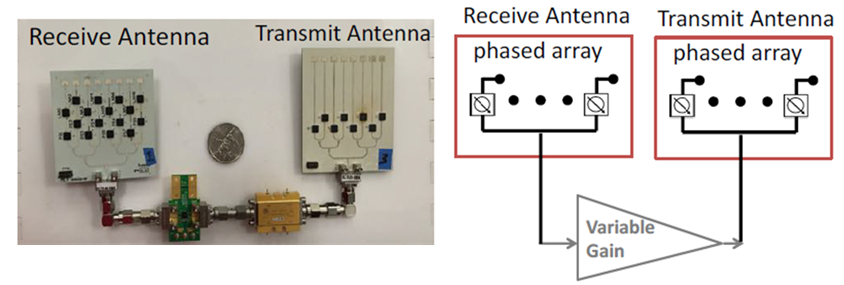}}
  \caption{A programmable mmWave mirror prototype and block diagram. (Reprinted, with permissions, from~\cite{Abari17mmNets}).}
  \label{fig.mmWaveMirror}
\end{figure}

The first straight forward conceptual architecture of RIS that can support a good set of the basic functionalities is simply a receive antenna array with its tunable beam-forming network  and a transmit antenna array with its independent tunable beam-forming network connected together, as shown in~Figure~\ref{fig.RISconceptarch}(a).  The receive antenna array with proper configuration of its beam-forming network captures energy of the incoming plane or spherical wave, then passes it to the transmit antenna array, which in turn through a proper setting of its beam-forming network, sends the energy towards the desired direction, either as a plane or spherical wave. An example of such RIS at a small scale can be found at~\cite{Abari17mmNets}. In~\cite{Abari17mmNets}, a simple programmable mmWave mirror is used to enable high quality untethered virtual reality application. The mirror prototype is shown in Figure~\ref{fig.mmWaveMirror}. The advantage of such an architecture is its conceptual simplicity and flexibility. However, the analog beam-forming network can become very complex and lossy quickly except for simple plane wave excitations, especially when the array size is scaled up. In addition, for large array and at low frequencies, the room needed for the two separate arrays also become a problem.

\begin{figure}
  \center{\includegraphics[width=.93\linewidth]{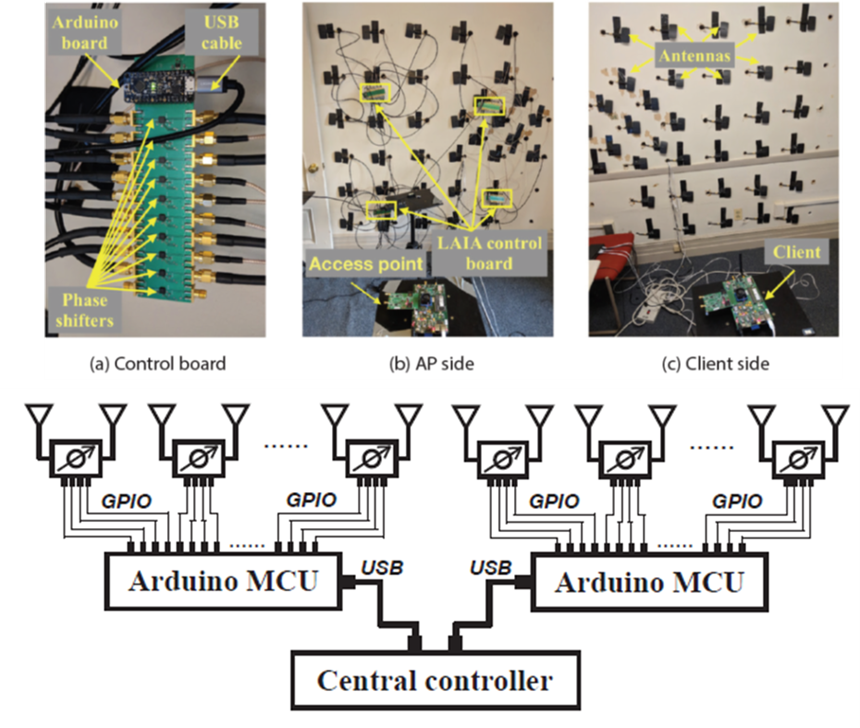}}
  \caption{Prototype and architecture of a large array of inexpensive antennas. (Reprinted, with permissions, from~\cite{ZLi17NSDI})}
  \label{fig.NSDILargeArray}
\end{figure}

The second equally straightforward conceptual architecture is similar to the first, except that it eliminates the analog beam-forming networks and opts for a one-to-one connection between the elements of the receive antenna array and that of the transmit antenna array with tunable phase shifters, as illustrated in Figure~\ref{fig.RISconceptarch}(b). This is essentially the architecture of a transmitarray antenna. An example of such RIS implementation can be found at~\cite{ZLi17NSDI}. In~\cite{ZLi17NSDI}, a prototype of 36-element array of inexpensive antenna is built with programmable phase shifters to demonstrate the concept of programming the radio environment. The prototype and architecture of this design is shown in Figure~\ref{fig.NSDILargeArray}. Besides the conceptual simplicity and flexibility, this architecture is also simple to implement. In addition, since the phase response of each individual pair of receive-transmit element can be independently controlled, it can more easily deal with complex wavefronts as discussed in previous sub-section. It is also the easiest to model mathematically. However, its major limitation is that it cannot support well functionalities that require non-local interactions, such as anomalous reflection as discussed in~\cite{DiRenzo20JSAC}.

\begin{figure}
  \center{\includegraphics[width=.98\linewidth]{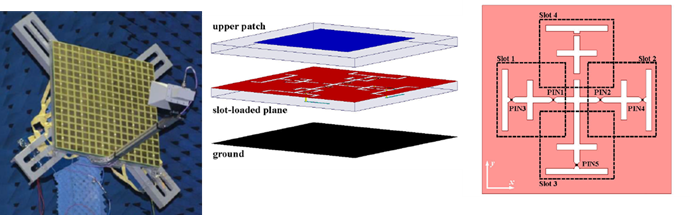}}
  \caption{Photo of a fabricated 2.3GHz RIS prototype and the design of the patch element with phase control. (\copyright 2020, IEEE. Reprinted, with permissions, from~\cite{LDai20Access})}
  \label{fig.RISprototypeLDai}
\end{figure}

A further refinement of the second conceptual implementation is to reuse the antenna elements for receive and transmit. It reduces the implementation to simply an antenna array  with tunable reflecting load. The received signal at each element is essentially reflected back though the same element with a variable phase controlled by its corresponding tunable reactive load, as illustrated in Figure~\ref{fig.RISconceptarch}(c). Of course, there are more ways to modify the response of an antenna element than just changing its load impedance. The element itself can directly be tuned to modify its properties, such as its resonant frequency, or resonant mode, or current distribution, etc., to effect a change in its response. Thus, we arrive at the architecture illustrated in Figure~\ref{fig.RISconceptarch}(d). This is essentially the architecture of reflectarray antenna. An example of such RIS implementation can be found at~\cite{LDai20Access}. In~\cite{LDai20Access}, 256-element RIS prototypes for 2.3GHz and 28.5GHz were developed with each element supporting 2-bit phase shift implemented using PIN diodes. One of the prototype design is shown in Figure~\ref{fig.RISprototypeLDai}. The PIN state configurations modify the RF current distribution, resulting in a change in the phase of the reflected waves.

\begin{figure}
  \center{\includegraphics[width=.98\linewidth]{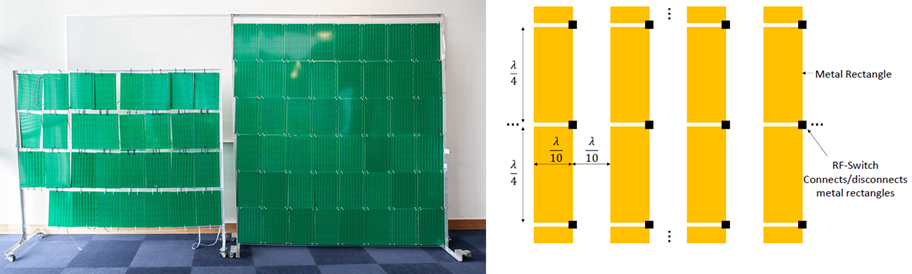}}
  \caption{RFocus prototype photo and schematic. Adapted from~\cite{Arun20RFocus}.}
  \label{fig.RFocus}
\end{figure}

\begin{figure}[t]
  \center{\includegraphics[width=.95\linewidth]{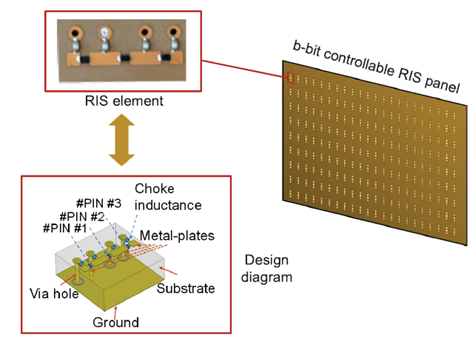}}
  \caption{Schematic structure of $b$-bit encoded RIS. (\copyright 2020, IEEE. Reprinted, with permissions, from~\cite{BDi20JSAC})}
  \label{fig.bBitRISstructure}
\end{figure}

So far, the antenna elements are still traditional in the sense that the element size and spacing is about a half of a wavelength of the incident waves, so the coupling between the elements is sufficiently weak to be ignored, and each element’s effect can be modeled independently from its neighbors.  As the element size and spacing shrink to far below the wavelength, the implementation becomes more a meta-surface~\cite{FYang19SurfEMBook}, where  a electrically thin surface with large footprint is composed of unit cells much smaller than a wavelength closed packed together with a spacing much less than a wavelength. Examples of such RIS implementation can be found in~\cite{Arun20RFocus,BDi20JSAC}. In~\cite{Arun20RFocus}, a very large low cost RIS is prototyped with 3200 elements, each of size of $\lambda/4$ and spacing of $\lambda/10$, as shown in Figure~\ref{fig.RFocus}. Each element has only 1 RF switch to control its state as either on or off. In~\cite{BDi20JSAC}, a design of meta-surface RIS with multi-bit encoding is proposed and studied theoretically. The schematic of the design is shown in Figure~\ref{fig.bBitRISstructure}.  The elements are claimed to be much smaller than a wavelength, however, there is no detailed dimension information.

\begin{figure}
  \center{\includegraphics[width=.95\linewidth]{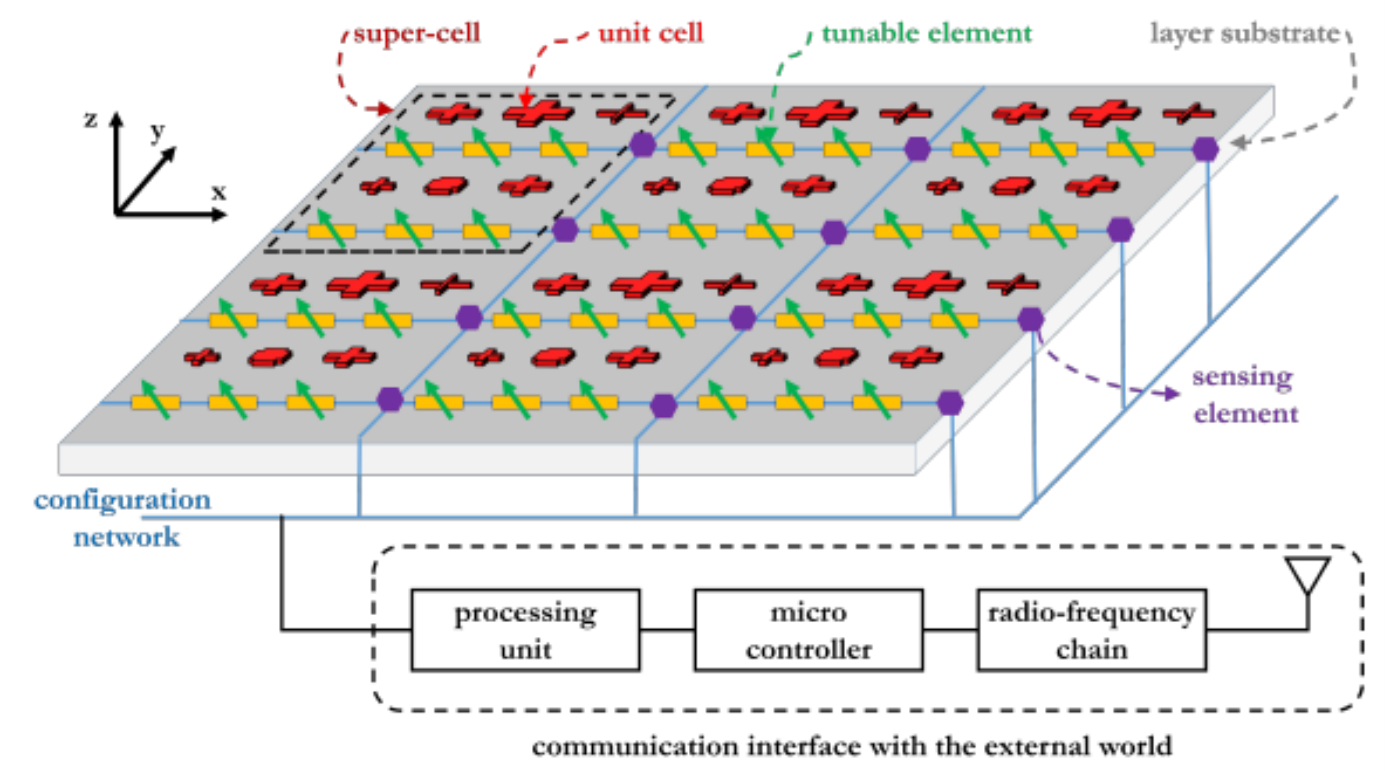}}
  \caption{Conceptual structure of an RIS based on meta-surface. (\copyright 2020, IEEE. Reprinted, with permissions, from~\cite{DiRenzo20JSAC}).}
  \label{fig.RISConceptArchMS}
\end{figure}

Due to the close spacing of the elements or unit cells in the meta-surface, there are strong interactions among them, and they can no longer be thought of or modeled as independently controlled scattering/reflecting elements. A more vigorous treatment is to abstract the surface as an infinitely thin boundary that creates discontinuity in the EM fields~\cite{Kuester03TAP,Holloway05TEC}. A conceptual generic model of such meta-surfaces is proposed in~\cite{DiRenzo20JSAC} and reproduced here in Figure~\ref{fig.RISConceptArchMS}.

In this conceptual model, the RIS is composed of multiple layers of periodic structures called super-cells. Each super-cell is composed of several unit cells of various shapes and sizes that act as passive scatters. In general, the unit cell is much smaller than a wavelength ($\sim$5 to 10 times smaller) and the spacing between the unit cell is of the same order. There exists strong interaction between the unit cells and their response must be modeled together instead of individually. A configuration network is necessary to control and program the tunable electronic circuits, e.g., PIN diodes or varactors, to achieve desired response from the surface. The model also includes means to communicate with the external world, so that it can be controlled and programmed remotely, or potentially provide sensing information collected by optional on-board sensors. \cite{DiRenzo20JSAC} contains a lengthy section detailing design methodology for specific functionalities based on this conceptual model, which interested readers are encouraged to read further.

\subsection{Brief Overview of Transmitarray and Reflectarray Antennas}
\begin{figure}
    \centering
    \subfigure[]  
    {   \centering
        \includegraphics[width=.30\linewidth]{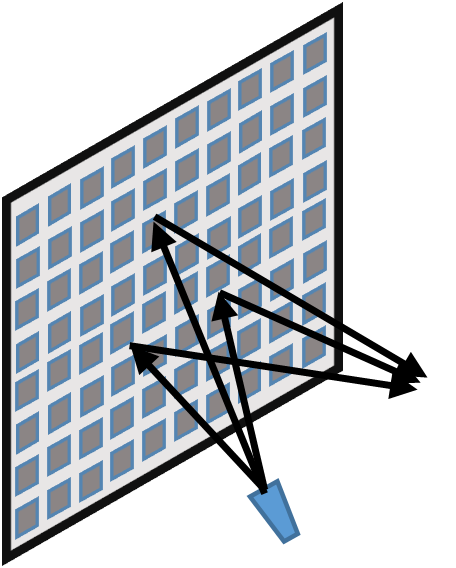}
    }
    \subfigure[]
    {   \centering
        \includegraphics[width=.45\linewidth]{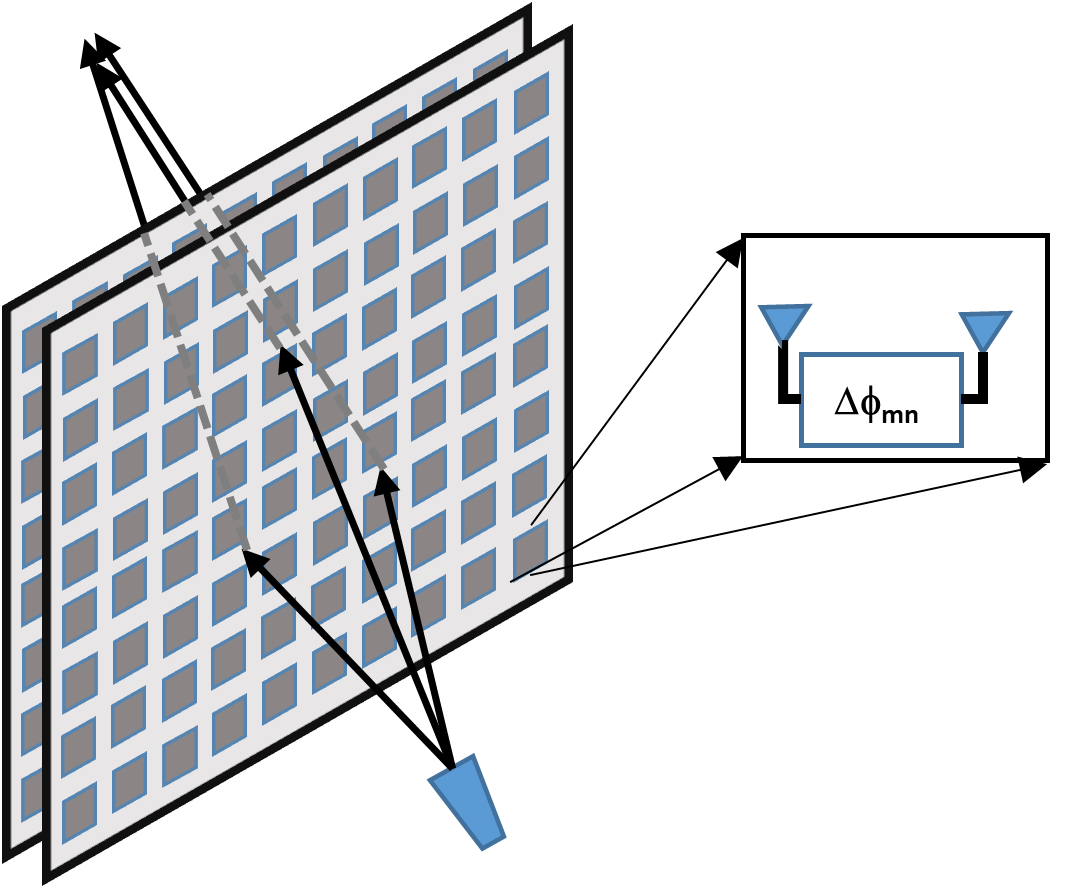}
    }
    \caption{Illustration of (a) reflectarray antenna and (b) transmitarray antenna geometries.}
    \label{fig.AntGeometry}
\end{figure}

The concept of using a planar or conformal surface with properties that can be manipulated to dynamically influence the scattered fields can be traced to transmitarray and reflectarray antennas.   Traditional reflector antennas utilized a shaped reflector surface with a fixed transmit feed location to produce antenna systems with large gains, and achieved beam steering by physically orienting the shaped reflector in the desired direction.  Reflectarray antennas maintained the concept of a  feed antenna, but now introduce various methods to synthesize a phase response on the typically planar reflector surface such that the scattered fields can be directed in the desired direction.  In a similar vein, transmitarray antennas, sometimes referred to as array lens, utilize a synthesized phase response to focus the impinging fields on the other side of the source through the surface. The reflectarray and transmitarray concepts are illustrated in Figure~\ref{fig.AntGeometry}(a)  and Figure~\ref{fig.AntGeometry}(b), respectively~\cite{NayeriReflectAntBook}.



This section of the paper attempts to elucidate the hardware needs, reconfigurable nature, complexities, and directions of RIS through the learned experiences of transmitarray and reflectarray antenna research community.  Of course, RIS will require a leap forward beyond current designs to dramatically upscale the hardware as to create very large, low profile, invisible, and conformal surfaces that must be controlled by an intelligent autonomous agent in the communication system.

The phase distribution over the reflectarray aperture must be dynamically changed in order to realize beam scanning functionality.  For the case shown in Figure~\ref{fig.RISbasicfun}(a), the phase distribution on each element of the reflectarray can be given by
\begin{equation}
\phi_{m,n}=k_0 \vec{R}_{m,n}-\Delta\phi_{m,n}.
\label{eq.phimn}
\end{equation}
where $\vec{R}_{m,n}$ represents the phase due to the spatial phase difference between the $(m,n)$th element and feed location.  While $\Delta\phi_{m,n}$ represents the phase shift introduced between the incident and scattered field on the $(m,n)$th element.  $k_0$ is the free space wave number.

From~(\ref{eq.phimn}) there are two methods to induce a dynamic phase shift over the aperture.  One method is by modifying the first term in~(\ref{eq.phimn}) by the so-called feed-tuning techniques.  As an example, this can be achieved by switching through multiple feeds or moving the feed relative to the reflect array.  The second method is by modifying the second term in (\ref{eq.phimn}) by the so-called aperture phase tuning techniques.  A combination of both methods is also possible.   For this paper, more detail about the second method is provided as more research efforts are focused in this area with novel enabling technologies.

\begin{figure}
\centering
\subfigure[]
  {\centering
  \includegraphics[width=.28\linewidth]{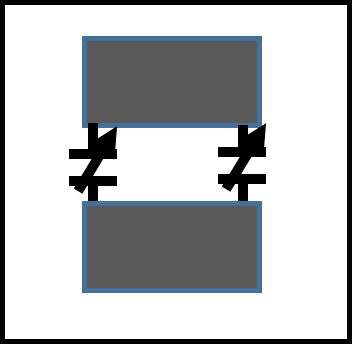}
  \label{fig.phasetuning.a}
  }
\subfigure[] {
  \centering
  \includegraphics[width=.28\linewidth]{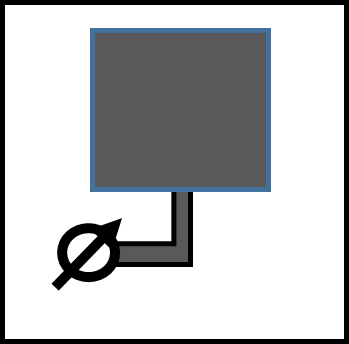}
  \label{fig.phasetuning.b}
  }
\subfigure[] {
  \centering
  \includegraphics[width=.30\linewidth]{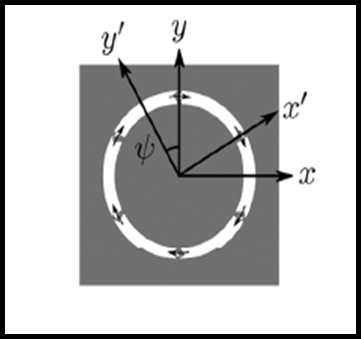}
  \label{fig.phasetuning.c}
  }
\caption{Reconfigurable phase tuning for reflectarray elements, (a) tunable resonator (or size), (b) tunable delay line, (c) element rotation.  (\copyright 2014, IEEE. Reprinted, with permissions, from~\cite{Hum14TAP}).}
\label{fig.phasetuning}
\end{figure}

Normally, reflectarray antennas are constructed from lumped elements not unlike the driven radiating elements found in some traditional antennas.  As an example, patch or microstrip type antennas on printed-circuit-board (PCB) material are commonly used. To enable these elements to dynamically change the phase over the aperture, the characteristics of resonance of the element must be changed or a variable phase shift must be added per element~\cite{Hum14TAP}.  As an example, three common elements are shown in Figure~\ref{fig.phasetuning}.

%

The element rotation method shown in Figure~\ref{fig.phasetuning}(c) is used for circularly polarized fields, and can be realized with rotation of the element via micro-actuators or electronically with a spiraphase phase shifter.

\begin{figure}
  \center{\includegraphics[width=.3\linewidth]{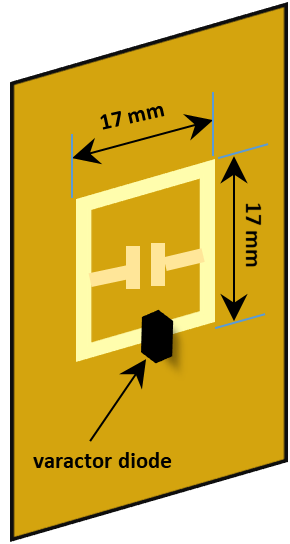}}
  \caption{ Unit cell for a metamaterial based reflectarray elements.  (\copyright 2010, IEEE. Reprinted, with permissions, from~\cite{Hand10AntPropLett})}
  \label{fig.unitcell}
\end{figure}

While Figure~\ref{fig.phasetuning} shows some examples of reflectarray unit cells, of course many other designs are possible.  One area of growing interest is to use engineered materials, or metamaterial, with unit features that are significantly smaller than the wavelength of operation. An example of this is given in Figure~\ref{fig.unitcell} where the meta-particle trace is approximately $\lambda_0/8$ in size.

\subsection{Typical Technologies to Achieve Reconfigurable Transmit and Reflectarray}
The previous section provided some examples of the elements that can be used to construct the reflectarray.  In this section some of the technology needed to implement the reconfiguration of the reflectarray or transmitarray is given in Table~\ref{tab.RRAnTA}, which is recreated from~\cite{Hum14TAP}, provides some examples of the technologies that can be used with particular designs to provide the tuning of the phase across the aperture of the array.  As well an assessment of the level of the technology parameters is given with associated references for further study.

\begin{table*}[t]
\centering
\begin{tabular}{|c|c|c|c|c|c|c|c|c|c|c|}
\hline
{\bf Type} & {\bf Technology}
& \rotatebox{90}{\bf Reliability - Maturity}
& \rotatebox{90}{\bf Integration}
& \rotatebox{90}{\bf D/A Control}
& \rotatebox{90}{\bf Complexity (Cost)}
& \rotatebox{90}{\bf Loss (Microwave/THz)}
& \rotatebox{90}{\bf Bias Power Consumption}
& \rotatebox{90}{\bf Linearity}
& \rotatebox{90}{\bf Switching Time}
& \rotatebox{90}{\bf References}\\
\hline
\multirow{3}{*}{\bf Lumped Elements}
        & PIN Diodes        &+  & - & D & + & -/-& - & O& + &	\cite{Hum07TAP} \\  \cline{2-11}
        & Varactor Diodes   &+  & - & D & + & -/-& + & -& + &	\cite{Petosa12AntMag} \\ \cline{2-11}
        & RF - MEMS         &O  & + & D & + & +/O& + & +& O &	\cite{Rajagopalan08TAP}\\
\hline
	{\bf Hybrid}  & Ferro-electric Thin Films	& O & + & A& O & O/- &	+ &	O &	+ &	\cite{Rajagopalan08TAP}\\
\hline
\multirow{3}{*}{\bf Tunable Materials}
        & Liquid Crystal  &O &O &	A &	O &	-/+ & O & O	&   & \cite{Long11AntPropLett}\\  \cline{2-11}
        & Graphene          &- &+ & A &	O &	-/+ & +	& - & +	& \cite{Carrasco13AntPropLett}  \\ \cline{2-11}
        & Photo-Conductive  &O & - &A	&O & -/- &	- &	- & + &	\cite{Chaharmir06TAP}\\
\hline
\multirow{2}{*}{\bf Mechanical}
        & Fluidic         & O & - & A & O & O/+ &	+ &	O &	- &	\cite{Rajagopalan08TAP}\\  \cline{2-11}
        & Micromotors       & - & - & A & - & + & O	& +	& - &	\cite{Rajagopalan08TAP} \\
\hline
\end{tabular}
\caption{Selected Technologies for the Implementation of reconfigurable reflectarray and transmitarray and Qualitative Assessment of a Few Related Parameters (‘+’,`O’,`-' Refer to Good, Neutral, and Poor, Respectively.  Reproduced from~\cite{Hum14TAP}.} \label{tab.RRAnTA}
\end{table*}

While there are a number of technologies available to provide phase tuning in the transmit or reflect arrays the best choice must be part of the overall antenna and system design.  As well as the frequency increases into the THz band, as envisioned for 6G, some of the solutions, such as varactors become too large and reconfigurability must be provided by control over the material properties.
\begin{figure}
  \center{\includegraphics[width=.9\linewidth]{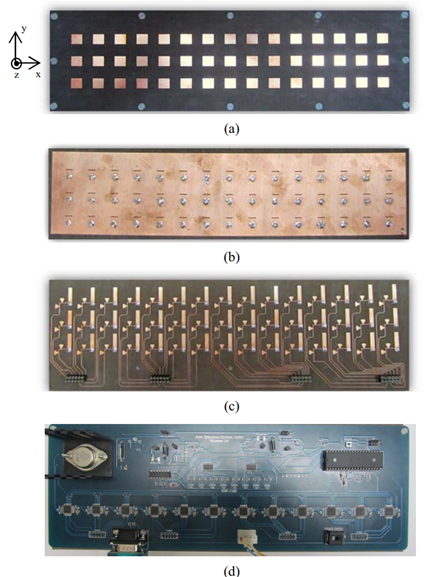}}
  \caption{Prototype of a 3$\times$15 element x-band reconfigurable reflectarray with 4 major components (a) patch elements (b) ground plane with slots feeds (c) varactor loaded lines with biasing traces, and (d) DAC  board that sets bias control line voltages.  (\copyright 2012, IEEE. Reprinted, with permissions, from~\cite{Venneri12EUCAP})}
  \label{fig.xBandProto}
\end{figure}

An example of a reconfigurable reflectarray prototype is shown in Figure~\ref{fig.xBandProto},   The major components of a typical reconfigurable array are shown and it is worth reviewing in more detail.  The top layer consists of patch elements on PCB, followed by the ground plane reflector with feed slots, followed by the PCB board with varactor loaded lines and corresponding bias lines, and finally the digital-to-analog converter (DAC) board that interfaces between the microcontroller and varactor bias line.

\begin{figure}
\centering
\subfigure[]{
  \centering
  \includegraphics[width=.40\linewidth]{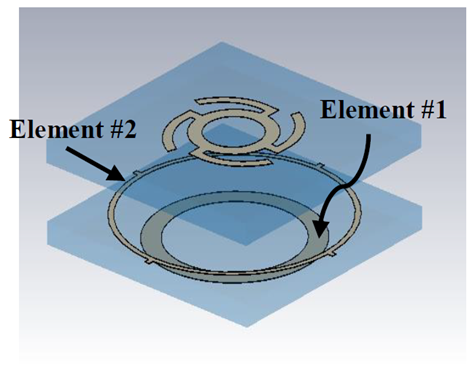}
  \label{fig.UnitCellProto.a}
}
\subfigure[]{
  \centering
  \includegraphics[width=.52\linewidth]{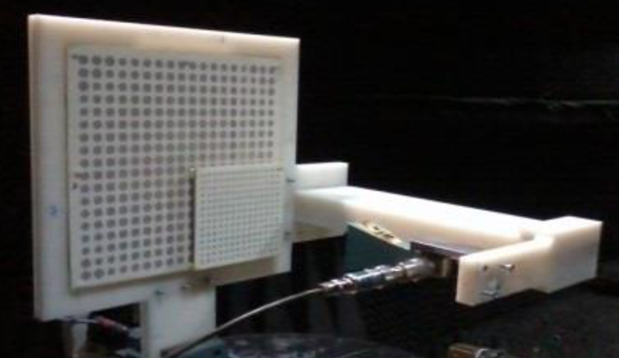}
  \label{fig.UnitCellProto.b}
}
\caption{(a) Unit cell for a frequency selective surface based dual band design and (b) fabricated prototype. (Reprinted, with permissions, From~\cite{Derafshi16JMOEA})}
\label{fig.UnitCellProto}
\end{figure}

Shown in Figure~\ref{fig.UnitCellProto}(a) is a reflectarray comprised of frequency selective surface backed unit cell design.  This dual band design allows for simultaneous operation in the X band (8-12 GHz) and Ku band (12-18 GHz).  From Figure~\ref{fig.UnitCellProto}(b), it can be seen that the reflectarray is comprised of two distinct surfaces and the use of the frequency-selective surface (FSS) for these surfaces allows for the isolation to be maintained between the bands.  Novel designs like this are further enhancing the frequency range and bandwidth of the reflectarray antennas, and are a research area for future explorations.
%

Similar in nature to the reflectarray is the transmitarray or array lens. The transmit array combines the principles of phased arrays and lens to allow the impinging wave to travel through the surface.  As such, the source is always on the opposite side of the desired user location.  One way to conceptualize the transmit array is illustrated in Figure~\ref{fig.AntGeometry}(b), where the surface can be thought of being comprised of two antennas, one receiving the field from the feed location and another transmitting the signal after a phase adjustment. This design approach is known as the guided wave method.  Another is the so-called layered-scattered approach where periodic structures such as slots, rings, or patches, in one or multiple layers, change the electrical behavior of a surface to allow transmission.  In any transmitarray design it is critical that the surface features allow for the transmission amplitude of the field to remain as close to unity while at the same time allowing phase shifts through 360.  In this way the losses are minimized, and the lensing capabilities are maximized.

\begin{figure}
\centering
\subfigure[]{
  \centering
  \includegraphics[width=.51\linewidth]{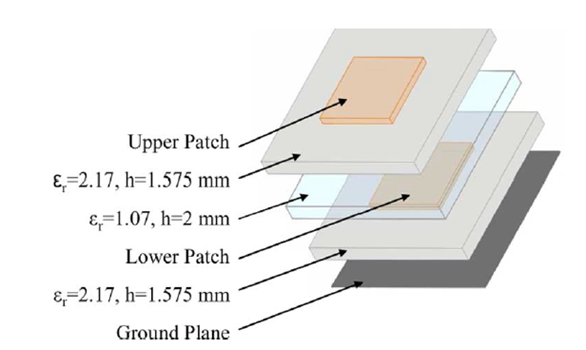}
  \label{fig.TxArrayAssembly.a}
}
\subfigure[]{
  \centering
  \includegraphics[width=.40\linewidth]{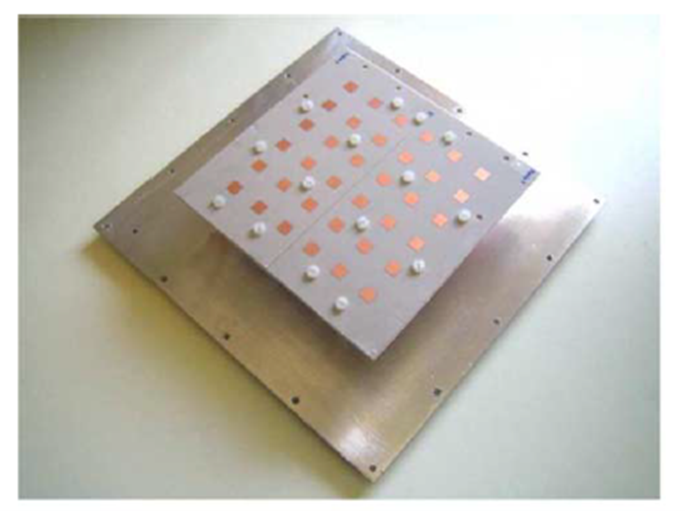}
  \label{fig.TxArrayAssembly.b}
}
\caption{Transmit array unit cell (a) and transmit array complete assembly without feed (b).  (\copyright 2010, IEEE. Reprinted, with permissions, from~\cite{Padilla10TAP})}
\label{fig.TxArrayAssembly}
\end{figure}

An example of the transmit array unit cell can be found in Figure~\ref{fig.TxArrayAssembly}(a), and it shows the stack-up of a multi-layer patch element designed to operate in the Ku band~\cite{Padilla10TAP}.  Figure~\ref{fig.TxArrayAssembly}(b) shows the prototype of the transmit array utilizing 36 patch elements, and what is not seen is the distribution network and varactors that provide for a reconfigurable phase shifter design to control the transmit array response.
%
%

\begin{figure}
  \center{\includegraphics[width=.7\linewidth]{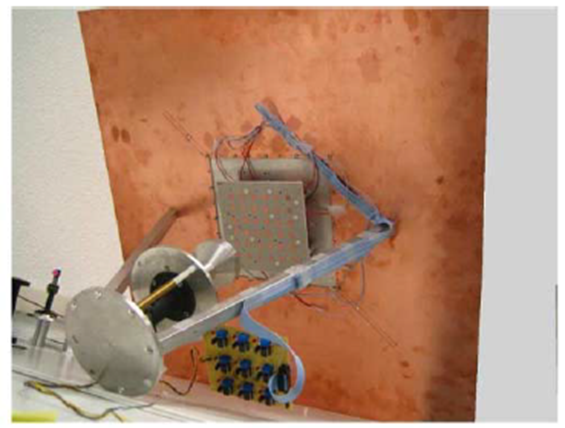}}
  \caption{Complete transmit array prototype with feed antenna and reconfigurable surface.  (\copyright 2010, IEEE. Reprinted, with permissions, from~\cite{Padilla10TAP})}
  \label{fig.TxArrayRISproto}
\end{figure}

Finally, a complete transmit array prototype is shown in Figure~\ref{fig.TxArrayRISproto}.  The measurements from the protype compare very well to the simulation results.  A couple of scenarios were measured.  First, in the case when the lens array allows field to pass through and does not change the radiation pattern, but the system corrects for any phase mismatches.  Second, then when the lens array changes the radiation pattern in one of the main axes, by applying a 9-degree tilt.  The last scenario demonstrates the ability of the transmitarray to steer beams dynamically.

There is a growing interest in transmitarray and reflectarray technology. Some of the research areas being explored are bandwidth enhancement, improved linearity, active reflectarray that integrate amplifiers in the design, and THz operation.  As well, electrically large arrays ($100$'s $\lambda$) have already been built in the Ka band, which gives hope that RIS-scale surfaces with transmit and reflectarray design are feasible.

\subsection{Hardware Design Challenges and Potential Research Directions}
As discussed earlier, the basic RF functionalities expected from an RIS are not much different from a reflectarray or transmitarray. Design concepts from the more mature reflectarray/transmitarray are readily applicable to many RIS designs. However, RIS does have additional challenges as it is envisioned to be much larger in size electrically and be used in much more complex and dynamic environment. The wavefront that it has to handle is much less likely to be that of a simple plane wave or spherical wave due to multi-path propagation, unlike the case for reflectarry/transmitarray design. The RIS may be expected to implement multiple basic functionalities with a larger number of goals simultaneously. In addition, due to the large size envisioned, it has to support wideband/multi-band operation, further complicating the design requirement.

In general, the synthesis of the meta-surface from desired EM properties is still an open problem, especially for complex requirements likely to arise for RIS. Hence it is a rich field for innovation.

The trade-off between size, cost and complexity is a challenge that needs to be tackled before application of RIS in any meaningful scale can occur. For example, trying to ram many features into a smaller size will likely increase the hardware design complexity and hence cost. By keeping it large and simple, so that it can be divided into multiple sub-arrays with each supporting only limited functionality, may allow lower material cost, but may significantly increase the deployment cost. Innovative ways to fundamentally change such trade-offs and limitations are needed for the success of RIS application in the real world. For example, the integration of RIS with building materials such as window glasses as demonstrated in~\cite{DOCOMO20TrialMetaSurf} is a very interesting direction.

Finally, with large scale RIS, the end points of communication will likely be in the near-field of the RIS, especially in indoor applications. Since most models and our own intuitions about antennas are based on far-field assumptions, there is still much to explore and understand. Potentially some phenomena unique to near-field could be leveraged to develop new applications or significantly enhance the performance of existing applications.

%% file: 05-AIML.tex
'\section{AI/ML for RIS Solution Design}\label{sec.aiml}

RISs have the potential to revolutionize the design of wireless networks, particularly when combined and integrated with other 6G candidate technologies such as AI-empowered wireless networks and terahertz communications.

Recent years have seen an overwhelming interest in using AI for the design and optimization of wireless networks. It has been gradually introduced in 5G cellular networks, more as isolated applications, and has been envisioned as a key enabling technology for 6G~\cite{DoCoMo2020WhitePaper,ali20206g,park2020extreme,Tariq20WirelessComm}.

In RIS-aided communication applications, there are more appealing reasons and motivations to apply AI/ML as part of the total solution, especially to control and optimize the re-configurable panels and elements. The massive use of meta-surfaces, reconfigurable reflectarrays, reconfigurable large-intelligent surfaces, provides a large number of degrees of freedom whose optimization presents a huge challenge to traditional analytical and numerical approach and entails a large computational complexity. In this section, we would like to discuss the motivations of AI/ML based approach in Section~\ref{sec.aiml.motiv}, followed by a detailed discussion of  recent research works and experiments in Section~\ref{sec.aiml.design}. We conclude with a summary of the strength and challenges of AI/ML based approaches for RIS-aided applications in Section~\ref{sec.aiml.diss}.

\subsection{AI/ML for RIS solution - motivations}\label{sec.aiml.motiv}

The growing interests and research of applying AI/ML for RIS-aided applications are motivated by two forces - ``push'' and ``pull''.

{\bf ``Push'' by growing problem complexity and demanding requirements}. Firstly, it is apparent that RIS-assisted networks/systems, such as communication, sensing, wireless charging, etc. are complex and challenging to design. This originates from the large number of parameters to be optimized based on the contextual information, and real-time decisions to be made every time when the network conditions change, e.g., channel condition changes, positions of the users change, etc.

Traditional or analytical solutions build upon communication theories, mathematical models, and optimization algorithms, which have been illustrated in earlier sections. While solid and very successful, they face new challenges in supporting RIS-assisted systems and applications.

To realize the optimal control of the RIS panels, one of the first and fundamental steps, in traditional approaches, is to “sense” and acquire channel state information (CSI). As pointed out in~\cite{yuan2020RIS},  the CSI acquisition problem in an RIS-assisted system has its unique challenges. High-dimensionality of the channel space introduced by the large number of RIS elements implies higher pilot training overhead (in both the number of pilot symbols required and the associated processing power) as well as longer delay. The passiveness of RIS poses extra challenges for signal processing algorithms. Furthermore, as in any model-based approach, the quality of estimated or reconstructed channel heavily depends on the accuracy of the underlying model being assumed. Currently the channel model of an RIS-assisted MIMO system has not yet been well understood. On top of that, even allowing for a well suited model, the required accuracy of channel estimation is very much application dependent. Current analytical solutions have no well established mechanism to address such a tradeoff between needed estimation accuracy and incurred algorithm processing overhead.

Secondly, as highlighted in~\cite{park2020extreme}, we are entering into a completely new space with “extreme” performance requirements in future communication applications, in terms of delay, reliability, throughput, etc., accompanied with higher frequency spectrum deployment. In traditional methods, as pointed out in~\cite{Jornod18WoWMoM}, the “sensed” channel status is mostly “after the fact” and the system can only react to what has already happened, which makes it extremely hard to combat any unfavorable or sudden change of propagation conditions, especially at high frequency bands.

Thirdly, at the system level, with or without RIS component, we will see a growing mixture of solutions including various cell layouts (macro cell, small cells, terrestrial as well as non-terrestrial networks), spectrum ranges (including millimeter wave (mmWave), terahertz, or even visible light communication), and a whole new dimension of channel dynamics when RIS is employed. It becomes very difficult, if not impossible, to completely rely on traditional model-based approaches to orchestrate and choose the optimal combination and configurations of these tools and/or options at various granularities.

{\bf ``Pull'' by the success of data driven knowledge learning}. We have witnessed enormous success of AI and ML in, e.g., computer vision, natural language processing (NLP), network  planning and control, etc., by learning the models that capture extremely complicated high-dimension relationship which were not available or feasible from existing theories and domain knowledge.

The traditional communication design, relying on theories and models, assumes minimum or zero knowledge of surrounding environment. As observed in~\cite{sheen2020digital}, the ``extreme'' performance requirements of many new applications that are anticipated demand a ``{\bf close fitness}'' between the radio control and resource management decisions and the underlying physical environment, e.g., its surroundings and user distributions/mobility etc. Conventional analytical approaches, while having a great strength leading us this far,   faces growing challenges and limitations to find such “close fitness” to the underlying environment and meet higher performance demands. 

\begin{figure}[t]
  \center{\includegraphics[width=.8\linewidth]{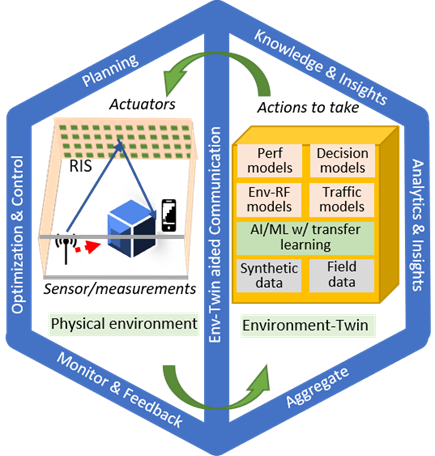}}
  \caption{Framework of Digital Twin for RIS-embedded environment.}
  \label{fig.DigitalTwin}
\end{figure}

This does not have to be the case anymore as argued in~\cite{sheen2020digital}. We are at the beginning of a new era of industrial transformation enabled by sensing, digitization, and connectivity. Combined with the growing power of  AI/ML algorithms, digital twins can be constructed for real-world entities, whether it be an organization, building, road, city, product, person, or process.  Such digital twins, for the first time, allows the communication system to self-learn the relationships among high dimensional factors and leverage the discovered knowledge of the underlying environment and network behaviors to make proper decisions at various granularity levels accordingly. The generalized digital-twin framework for RIS-embedded environment is depicted in Fig.~\ref{fig.DigitalTwin}.

\subsection{Research and experiments of AI/ML in RIS solution design}\label{sec.aiml.design}

Recently, there are many research publications that study and experiment with AI/ML based approaches for RIS-aided application solutions. In this paper, we organize them in the following three groups:

1)	ML to assist channel status acquisition

2)	ML to enable solutions without explicit channel estimation

3)	Hybrid solutions combining both model based and model free approach.

\subsubsection{ML to assist channel estimation or reconstruction}

To handle the challenges in channel estimation or reconstructions, there are researches and studies that apply a new, ML-based  approach, or try to learn the structure features of the channels to reduce the overhead of CSI acquisition.

Normally, ML-based approaches require data to train a model for supervised learning. For channel estimation, the input usually includes received pilot signals and the output is the predicted channel state. Typically, the training is done in a centralized manner as proposed in~\cite{Elbir20WCL} where the authors proposed ChannelNet, a twin convolutional neural network (CNN) architecture that is fed with the received pilot signals to estimate both direct and cascaded channels. In a multi-user scenario, each user has access to the CNN model to estimate its own channel. The labels for the direct channel estimation are generated with all the elements on the intelligent surface off, and the cascaded channel is estimated by transmitting pilot symbols when each of the RIS element is turned on one by one (then the least square method is used to estimate the cascaded channel). This centralized approach poses significant transmission burdens on the wireless links.  To overcome this, the authors in~\cite{elbir2020federated} developed a federated learning (FL)-based approach for channel estimation, in which a CNN is trained on local datasets of the user equipment (UE) without sending the data to the base station (BS). The same CNN model can be used to estimate both direct and cascaded channels in RIS-assisted massive MIMO network.  Simulation results showed such approach can reduce transmission overhead by 16 times compared to centralized approach.

In~\cite{yuan2020RIS} a “structure-learning based CSI acquisition” solution is summarized. The cascaded channel of an RIS-aided system usually exhibits strong structural features, such as sparsity and low-rankness, which can be exploited to reduce the overheads for the CSI acquisition. The sparsity of the cascaded channel is attributed to the deployment of large-scale antenna arrays, so that the RIS and the BS are able to distinguish EM waves from different directions with high resolution. The low-rankness of the cascaded channel arises from the effect of limited scattering in the user-to-RIS  and RIS-to-BS propagation environment. Moreover, signal sparsity can be artificially introduced to assist the cascaded channel estimation by controlling the on/off states of the RIS elements. With these structural information, the estimation of the cascaded channel can be done by utilizing advanced signal processing tools, such as compressed sensing, sparse matrix factorization, and low rank matrix recovery algorithms.

\subsubsection{ML to enable solutions without explicit channel estimation}

As discussed earlier, CSI acquisition is very challenging in RIS context. There are several interesting works advocating a more data- and learning-driven approach, which potentially reduces need of, or complements, the  conventional channel estimation or reconstruction.  We summarize and present these works at two levels described as follows.

Firstly, it is interesting to see a few groups of researchers proposing new visions and frameworks for AI empowered communication systems, sharing the same or similar observations and motivations as presented in the previous section. It is worth noting that these new frameworks present potential benefits and solutions in general, and not just limited to RIS-aided applications.

In~\cite{zeng2020environmentaware}, a vision of “environment-aware communication” is presented, which is enabled by “Channel Knowledge Map” (CKM), especially in addressing the practical challenges brought by the drastically increased channel dimensions and training overhead. The essential task of building a site-specific CKM is to establish an accurate mapping between the channel knowledge of interest and all potential transmitter/receiver locations in the target area/space, based on a finite amount of collected data tagged by locations and also possibly time stamps. This is feasible due to the strong spatial correlations of wireless links, together with the (quasi)-static nature of many aspects that impact wireless channels, such as the deterministic blockages, reflectors, and scatters. Once CKM is learned for a site, it can be utilized for training-free communications or CKM-assisted channel training (CKM as additional side-information). With CKM-enabled training-free communications, the channel knowledge predicted by the CKM is directly used for the system design and performance optimization, without requiring the conventional pilot-based channel training if the CKM is sufficiently accurate and/or when only coarse performance guarantee is needed.

In~\cite{DiRenzo19JWCN}, a “smart radio environmental” framework is enabled by the so called “environmental AI” together with reconfigurable meta-surface. The reconfigurable meta-surfaces may be equipped with embedded sensors that could allow them to sense the status of the environment, e.g., the channel states between them and the base stations, and between them and the mobile terminals. It also suggests the RIS control agents interact with the environment through their embedded sensors, and learn the optimal policy via reinforcement learning. One of the challenges using this approach is that, in wireless environments which are highly dynamic in nature, the system may not converge within the coherence time of the environment because of the well-known exploitation-exploration dilemma of reinforcement learning methods.

In~\cite{sheen2020digital}, a new digital twin based paradigm is proposed for RIS solution design, including a systematic framework of constructing the “environment-twin” (Env-Twin) to capture different aspects of an RIS-embedded environment and its behavior leveraging state-of-art ML algorithms. It proposes an Env-Twin consisting of a set of models that represent layered abstractions of the underlying environment and behaviors, each of which captures  the interaction and (ideally the) causal relationship, typically between States and Observations, or between States and Decisions.

On the next level, we would like to present some concrete design and experiments to study how ML can be used in designing solutions for RIS-embedded environment. In general, we group the approaches into three categories based on ML techniques leveraged, namely supervised learning, unsupervised learning,  and reinforcement learning.

\begin{itemize}
\item[a)]	Supervised learning: In a typical setting, RIS’s are passive and do not have sensing capabilities, which present challenges in optimal RIS operation given the massive number of elements on the RIS. In~\cite{Taha19GC}, the authors developed an RIS architecture that introduced only a small number of active RIS elements that are connected to the baseband, thus with sensing capabilities. The solution exploited deep learning (DL) techniques to predict optimal RIS reflection matrices using knowledge of the channels seen only at those active RIS elements. Through simulation validation, the proposed approach achieved near-optimal data rates with relatively small training overhead, and without any knowledge of the RIS geometry. In another research study~\cite{Liaskos19SPAWC}, the authors considered a multi-RIS-wall communication environment and regarded wireless propagation as a deep neural network (DNN) and designed a more  interpretable neural network (NN) architecture, where each RIS surface is represented as one NN layer, number of neurons are based on the number of elements on the RIS and their cross-interactions are represented as links between neurons. After training from data, the wireless network learns the propagation basics of the RIS and configures the elements to the optimal setting.

In~\cite{CHuang19SPAWC}, the authors proposed a DL-based approach to learn the mapping between the measured coordinate information at a user location and the configuration of each RIS element that maximizes the user’s received signal strength in an offline phase utilizing a preconstructed fingerprint database in training the DNN. In the online phase, the trained DNN model is fed with the measured coordinates of the target user location to output the optimal phase configuration focusing on the intended location.

In~\cite{sheen2020digital}, the authors introduced a different paradigm in predicting the optimal reflection beamforming vector for a receiver location, in which no explicit channel estimation effort is required, thus reducing the costly overhead in acquiring such information. The proposed solution learnt the mapping function between the receiver location attributes with any intended RIS configuration, and the corresponding achievable rate directly via a properly constructed DNN architecture. The authors designed their so-called Env-Twin DL neural network architecture leveraging domain knowledge for input feature representation and ML techniques. Simulation results showed that such approach can converge to near-optimal data rates using less than 2\% of the total number of possible receiver locations in the environment during training phase.

Although with good simulation results, supervised learning comes with the cost of collecting large amount of labeled training samples in advance. These training samples are usually very difficult and time-consuming to obtain.

\item[b)]
 Unsupervised learning: To overcome the dependency of collecting large amount of labeled data in advance as usually required in the supervised learning setting for model training, unsupervised learning is an appealing approach. In~\cite{JGao20CommLett}, the authors proposed an unsupervised leaning approach leveraging DL technique for passive beamforming design in RIS-assisted communication network. The proposed method assumes channel state information can be acquired and constructs the input using the product of the channels of the link between RIS and transmitter and the reflecting link between RIS and user, and the channel of the direct link. The output of the DL model is the predicted phase shift setting. As the approach is unsupervised, the authors defined the loss function as the negative of the objective function for maximizing the transmission rate. Simulation results showed the proposed method can achieve decent performance compared to the baseline semi-definite relaxation (SDR) approach while significantly reduced the system running time.

\item[c)]	Reinforcement learning: Given the overhead associated with supervised learning and recent success in deep reinforcement learning (DRL)~\cite{KYang20InfoCom} on resolving complicated optimization problems in wireless communication networks, some researchers proposed solutions leveraging DRL-based techniques. In~\cite{CHuang20JSAC}, the authors introduced an actor-critic DRL-based approach to study the joint design of transmit beamforming matrix at the base station and the phase shift matrix at the RIS together for multiuser MISO system. The design assumes direct transmissions between the BS and the users are totally blocked, and the channel matrix from the BS to the reflecting RIS and the channel vector from the RIS to all the users are perfectly known at both the BS and the RIS. The sum rate is utilized as the instant rewards to train the DRL based algorithm, and the transmit beamforming matrix and the phase shifts are jointly obtained (as the output of the DRL neural network) by gradually maximizing the sum rate through observing the reward and iteratively adjusting the parameters.

In~\cite{Taha20SPAWC}, the authors designed a potential standalone RIS solution also based on DRL. To acquire the channel information as input to the ML model, RIS is equipped with a few active elements to provide sensing capability. Similar to~\cite{CHuang20JSAC}, the target scenario assumes direct link between the transmitter and receiver is blocked, and LoS between the RIS and receiver. States used in the DRL solution are the normalized concatenated sampled channel (by the active elements) of each transmitter-receiver pair, actions are represented by each candidate interaction vector, and achievable rate at the receiver is used as the reward function to train a fully connected neural network. The DeepMIMO dataset is adopted to generate the labels based on the outdoor ray-tracing scenario ‘O1’. Simulation results based on accurate ray-tracing channels showed the proposed DRL approach can converge to near-optimal rates, close to the performance of the baseline approach which utilized  supervised learning.
\end{itemize}

\subsubsection{Hybrid solution combining both model based and model free approach}

Traditional analytical and optimization approaches rely on mathematically convenient models. On the other hand, ML-based approaches are data driven and less reliant on analytical system models. In~\cite{yuan2020RIS}, it is argued that the PHY-layer characteristics of wireless systems, such as link propagation and interference, are very well understood, despite the difficulty in modeling the entire complicated RIS system. Completely ignoring the model availability would negatively impact the algorithm efficiency. As such, it is desirable to design an integrated framework, where model-free learning and model-based optimization approaches complement each other and work better together. As an example, one possible solution was introduced using the actor-critic based deep reinforcement learning (DRL) framework.

Another interesting approach is presented in~\cite{Liaskos19SPAWC} “An Interpretable Neural Network for Configuring Programmable Wireless Environments”. Its problem setting is more complicated than previous ones, involving multiple panels so the optimization space is much bigger. It suggests constructing a “custom neural network” to adaptively configure RIS tiles for a set of users. The key idea is to represent each RIS tiles as nodes in a neural network, while the power distribution can be mapped to neural network links and the associated  weights. The weights are then optimized via custom feed-forward/back-propagation processes and are interpreted into RIS tile functionalities. The main difference between  this approach and other ML methods mentioned earlier, is that a model is assumed to exist connecting the tiles/nodes of the panels (panels are modeled as layers of the NN). In other words, the model tells how the power is eventually passed to layers of tiles if the configurations of each tile are known. Therefore, we regard this approach a hybrid one, which combines the known physical models with ML method (loss function and back-propagation), to find the optimal settings of each tile. As the weight of each link of the NN has physical meaning, the trained NN model is intuitive and interpretable. The challenges of this method are in two folds. Firstly, the physical model being used to connect the NN nodes needs to be in good quality in both modeling the physical properties, as well as reflecting the geo-setting (including  the relative locations of panels/tiles and mobile devices). Secondly, it is unclear how we can efficiently train models for all user locations and their combinations.

We summarize the above discussed AI/ML-based RIS solutions  and experiments in Table~\ref{tab.ALML4RIS}

\begin{table*}[t]
\centering
\begin{tabular}{|p{1.3cm}|p{2.4cm}|p{2cm}|p{1.8cm}|p{1.5cm}|p{5cm}|}
\hline
{\bf	Category } &	{\bf Scenario}/\;\;\;\;\; {\bf objective} & {\bf	Approach } &  {\bf ML Model} & {\bf Reference} & {\bf Key contribution}\\
\hline
(1) &
Channel estimation &
{Supervised learning (DL)}  &
Twin CNN  &
\cite{Elbir20WCL} &
Estimate both direct channel and cascaded channel at the same time.\\
\hline
(1) &
Channel estimation with lower transmission overhead  &
Supervised learning
(FL) &
CNN &
\cite{elbir2020federated} &
Reduce transmission overhead by 16X compared to centralized leaning approach.\\
\hline
(2) &
RIS-assisted network with few active elements  &
Supervised learning (DL) &
Fully connected neural network (FCNN) &
\cite{Taha19GC} &
Approach requires only a few active elements on the RIS, thus reduces beam training overhead.\\
\hline
(2) &
RIS-assisted indoor environment&
Supervised learning
(DL)	&
FCNN	&
\cite{CHuang19SPAWC} &
Learn the mapping between a user’s position and the optimal configuration of each RIS element with a fingerprinting database.\\
\hline
(2) &
RIS-assisted network with all passive elements &
Supervised learning (DL)	&
CNN &
\cite{sheen2020digital} &
Introduce an Env-Twin framework to model RIS-assisted environment exploiting location attributes without channel estimation overhead.\\
\hline
(2) &
RIS-assisted network with no labeled data 	&
Unsupervised learning
(DL)	&
FCNN &
\cite{JGao20CommLett} &
Introduce unsupervised learning approach with no need of collecting labeled training data in advance.\\
\hline
(2) &
RIS-assisted multi-user MISO system optimization&
Model-free actor-critic deep reinforcement learning (DL)	&
FCNN	&
\cite{CHuang20JSAC} &
Jointly optimize the transmit beamforming matrix and the phase shifts together.\\
\hline
(2) &
Standalone RIS operation &
Reinforcement learning (DL)	&
FCNN &	
\cite{Taha20SPAWC} &
Approach does not require any control from communication infrastructure. Validation showed achieving comparable performance as supervised learning.\\
\hline
(3) &
Multi-RIS-wall indoor environment &
Supervised learning (DL) &
FCNN &
\cite{Liaskos19SPAWC} &
Modeling wireless propagation environment with multiple RIS walls via an interpretable neural network architecture.\\
\hline
\end{tabular}
\caption{ Summary of existing ML-based RIS research works: (1) ML-based approach to assist channel state acquisition; (2) ML-based approach without explicit channel estimation; (3) Hybrid solution.} \label{tab.ALML4RIS}
\end{table*}

\subsection{Strength of AI/ML and challenges of their applications to RIS-aided systems}\label{sec.aiml.diss}
AI/ML provides powerful tools to gain additional knowledge and intelligence of surrounding environment that conventional system and solution fall short of achieving. For the first time, as pointed out in~\cite{sheen2020digital}, the communication system can learn the relationships among high dimensional factors, which is hard to capture using traditional analytical approach, and leverage the discovered knowledge of the underlying environment and network behaviors for performance and other optimizations.

For AI/ML in RIS-aided applications, we summarize the key advantages of a learning based approach as following.
\begin{itemize}
\item[1)] {\bf Low overhead and real-time control} of RIS elements. When sufficient prediction accuracy can be achieved by the learned models, it is possible not to rely on explicit channel estimation/reconstruction (i.e., training-free)   (\cite{sheen2020digital,zeng2020environmentaware}), thus reducing the overhead and latency from pilot-based training. This is important as we move towards more ultra-reliable low-latency communication (URLLC) driven applications, in conjunction with other types of traffic.
\item[2)] {\bf Predictive} control of RIS elements. As illustrated by several published works, with the help from ML the channel characteristics becomes predictable based on device locations and/or other fine-granularity context information. It is also anticipated that the predictions can be combined with other ML learned models for user/device trajectory/mobility and/or traffic distribution predictions. \cite{Subrt12IETComm} illustrated an approach where user distribution prediction is used to control the different panels’ ON/OFF setting. For high frequency band communication, the presence of a LoS path is an important assurance for high reliability communication needs. RIS can help “create” such LoS channel but only if we can predict the channel dynamics and configure RIS elements proactively. An interesting study in this regard can be seen in~\cite{koda2020distributed} where a split-learning architecture is proposed to enable “Vision Aided mmWave Received Power Prediction”.
\item[3)] 	{\bf Broad deployment and application scenarios.} With ML-based approach, the signal processing “burden” or requirements on communication devices and nodes are significantly reduced. For example, the RIS panel can be completely passive, allowing “standalone” deployment~\cite{Taha20SPAWC} without deep integration with existing communication systems. Note that~\cite{Taha20SPAWC} leverages few active elements on the RIS to perform sensing and channel estimation. It is also possible for RIS to work with devices that have limited or no signal processing capability. This will broaden applications of RIS technology for example in IoT applications, or in the wireless power transfer use case.
\item[4)] 	{\bf Beyond simple scenario.} Most of the researches so far are limited to simple problem settings (e.g., single-panel-single-user) and applying AI/ML for panel element control. Yet we can anticipate that ML models, which capture the interaction patterns of panel/elements with device locations etc., can be used for more complicated and challenging settings, e.g., multi-panel-multi-user scenarios. Also, they can help other stages of RIS solution design, e.g., optimizing multi-panel placements during the planning phase.
\end{itemize}
It is only natural that, while promising and with many potential advantages, the AI/ML approach for RIS solution design faces many challenges at the same time.  These challenges are also opportunities for future research directions.

Firstly, the effectiveness of the learned models depends on the availability, granularity, and the quality of the observation data, for both surrounding environment and the channel characteristics, and performance observed at the device side. The richer, more granular, and higher quality of the measurements, the better the quality of the trained models, enabling a greater number of more powerful applications.

Secondly, domain knowledge plays a critical role in the model input/output design, offering a clear  understanding of the underlying causal relationship and predictability, while considering the coherence time of the wireless environment. In some cases, due to the limited observation of state information, there is a high degree of uncertainty of the output. Reducing such uncertainty would require enrichment of the environment state representation, and/or further online exploration/exploitation as indicated in~\cite{DiRenzo19JWCN}.  Domain knowledge may also accelerate the convergence of the ML algorithms within the coherence time of the wireless channel.

Thirdly, tradeoff between site-specific model and generalizable knowledge learning needs to be better understood. It is true that the machine learned models are likely site-specific. However, it would be much desirable for such models to capture a more general knowledge of the underlying physics law (e.g., from the relative locations, distance, and objects in between etc.).  An interesting experiment described in~\cite{sheen2020digital} shows that it is possible to learn such knowledge from a small amount of labeled data. However, it remains a challenge to learn more transferable knowledge beyond simple problem settings, e.g., environment with multiple users or devices.

Lastly, we will face a whole suite of challenges to make the new AI based approach work in  real-world  experiments. For almost all the works we surveyed, the studies and experimental results are  based on numerical simulations, which are tractable first steps. However, it is generally recognized that the data from simulation are clean and the simulated behaviors are more deterministic. Simulation  also provides the convenience to generate the large quantities of the data required for machine learning. In a real-world environment these benefits would not be available. It would be critical to go beyond simulation based experiments to study the practical feasibility and effectiveness of the AI/ML-based RIS solutions. It is encouraging to see  in~\cite{Arun20RFocus}  a large-scale real-world design and implementation was introduced,  using non-conventional approach without explicit channel estimation for its control (the majority voting algorithm employed for RIS configuration exhibits an AI/ML flavor). Many other practical issues and challenges remain to be addressed for AI/ML-based RIS solutions, for example the distributed model training, updates, and transfers, just to name a few.

%% file: 06-Challenges.tex
\section{System Level Challenges and Future Directions}\label{sec.challenges}
The practical challenges presented in the preceding sections give rise to exciting new research directions in their own rights. Collectively they also contribute to \emph{system level} research opportunities to ensure a commercially viable deployment of RIS-embedded wireless network.

\begin{enumerate}
  \item An RIS-integrated B5G network must offer superior combined cost, power, performance and ease of deployment advantages against incumbent ones in order to be commercially competitive, especially for realistic yet complex multi-panel and/or multi-user scenarios. Factors such as surface design and manufacturing cost, overall computation complexity and associated power consumption (see Section~\ref{sec.antrf} for a comprehensive summary), and network control protocol overhead all become variables in the overall system design tradeoff.
      \begin{itemize}
      \item For instance, there is a fundamental tradeoff between the protocol overhead associated with environment state information acquisition and network capacity, especially for a system that employ a large number of nearly-passive distributed surfaces to ensure robust network coverage;
      \item Likewise, such massive deployments compound computational complexity required for optimal configurations for each surface, with associated power consumption that offsets the power saving promised by passive surfaces;
      \item Cost-effective nearly-passive hardware architectures such as severely quantized phase control and absence of sensing elements on the surface may demand additional capabilities in digital signal processing and optimization to compensate for performance degradation.
      \end{itemize}

      Therefore, in addition to analytical scaling laws governing fundamental system capacity limits, those that govern the performance-cost-power tradeoff in a real-world system require careful investigation, which in turn relies on realistic physically consistent surface interaction and propagation models to allow simulation-based comparative evaluation against legacy systems, especially for compact metasurfaces employing sub-wavelength spaced unit cells, and for the newly emerging near-field application scenarios enabled by large surfaces.

  \item The network architecture that is optimized for an RIS-embedded network may look quite different from ones for its RIS-free counterpart. For instance, nearly-passive RIS surfaces require frequent endpoints transmission to probe the channels between the surfaces and endpoints, and between different surfaces, in dynamic scenarios such as user mobility. In the case that the number of degrees of freedom of the surface-enabled environment exceeds the capability of the endpoint pairs to probe (which is conceivable given that extra and perhaps redundant number of degrees of freedom from artificial environment is sought after in the RIS paradigm), protocol overhead may dominate system performance. Either surfaces with sensing and transmission capability or additional `probing' endpoints maybe required. Both involve careful design of low-power protocol to transport sensed channel states from distributed measurement points to distributed or centralized processing entities.
  \item In contrast to conventional cellular deployments of today where endpoint transmitter power spectral density (PSD) masks suffice to ensure tolerable out-of-band emission between neighboring licensed bands belonging to different network operators in an overlapping environment, an RIS surface configuration optimized for one operator's band allocation may incur performance degradation to another operator's network that shares the same surface as part of its propagation environment, unless the configuration of the surface response can be made band-selective, or different networks deploy dedicated band-limited surfaces specific to their own band allocation. Such multi-operator system level compatibility requirements require further investigation.
  \item The optimization along the new environmental dimension enabled by RIS, and the latter's potential for joint sensing, learning and control of the environment of the dynamic ambient surroundings requires scalable solutions in an extremely high-dimensional space. Traditional numerical optimization techniques may fall short. Furthermore, low-cost and low-power surface architectures may render them and the associated propagation characteristics less amiable for parametric modeling. Model-free techniques such as AI offer unique advantage here. Effective means of capturing the underlying structural features of the environment remains a challenge for any AI techniques. See Section \ref{sec.aiml.diss} for AI-specific system level challenges.
  \item The question of ``killer'' application(s) for an RIS-embedded system remains open. One pressing research direction would be to identify which industrial verticals show the most significant benefit in terms of return on investment and total cost of ownership. It appears that applications that require ubiquitous spatial coverage with a modest degree of mobility, and/or those that derive added value from a synergistic combination of communication and near-field sensing, localization and power transfer functionalities, are likely to benefit the most, provided that the overall value to customers and network operators justifies the added cost associated with mass fabrication and protocol overhead.

\end{enumerate}

%
%
%
%
%
%

%% file: 07-Conclusions.tex
\section{Conclusions}\label{sec.con}
This paper provides a literature survey and outlook on practical design aspects of RIS, and system level challenges facing a commercially viable deployment of an RIS-integrated wireless network. In addition to summaries on salient recent works on system modeling and PHY design, comprehensive coverage are given for both hardware architecture designs and challenges, and unique opportunities such large-scale multi-purpose intelligent networks present to AI/ML techniques.